\documentclass[prd,superscriptaddress,nofootinbib,a4paper,preprintnumbers]{revtex4} 
\usepackage{graphicx}
\usepackage{dcolumn}
\usepackage{bm}
\usepackage{latexsym}
\usepackage{amsfonts}
\usepackage{amssymb}
\usepackage{amsmath}
\usepackage{bbold}
\usepackage{ulem}
\usepackage{url}
\usepackage{xcolor}
\usepackage[utf8]{inputenc} 
\usepackage{xspace}

\allowdisplaybreaks

\begin{document}

\title{Generalising the matter coupling in massive gravity: a search for new interactions}
\author{A. Emir G\"umr\"uk\c{c}\"uo\u{g}lu}
\email{emir.gumrukcuoglu@port.ac.uk}
\affiliation{Institute of Cosmology and Gravitation, University of Portsmouth\\ Dennis Sciama
Building, Portsmouth PO1 3FX, United Kingdom}

\author{Kazuya Koyama}
\email{kazuya.koyama@port.ac.uk}
\affiliation{Institute of Cosmology and Gravitation, University of Portsmouth\\ Dennis Sciama
Building, Portsmouth PO1 3FX, United Kingdom}

\date{\today}
\begin{abstract}
Massive gravity theory introduced by de Rham, Gabadadze, Tolley (dRGT) is restricted by several uniqueness theorems that protect the form of the potential and kinetic terms, as well as the matter coupling. These restrictions arise from the requirement that the degrees of freedom match the expectation from Poincar\'e representations of a spin--2 field. Any modification beyond the dRGT form is known to invalidate a constraint that the theory enjoys and revive a dangerous sixth mode. One loophole is to exploit the effective nature of the theory by pushing the sixth mode beyond the strong coupling scale without completely removing it. In this paper, we search for modifications to dRGT action 
by coupling the matter sector to an arbitrary metric constructed out of the already existing degrees of freedom in the dRGT action. We formulate the conditions that such an extension should satisfy in order to prevent the sixth mode from contaminating the effective theory. Our approach provides a new perspective for the ``composite coupling'' which emerges as the unique extension up to four-point interactions. 
\end{abstract}

\maketitle

\section{Introduction}
\label{sec:intro}
Modified gravity theories typically give rise to new degrees of freedom that contribute to gravitational interactions. These are either non-minimally coupled extra fields added by hand (e.g. scalar-tensor theory), or arise due to partial or complete breaking of the diffeomorphism symmetry (e.g. massive gravity). In the presence of these new forces, there is no longer a unique space-time measure. With the help of the new fields and their derivatives, one can define a new geometry by rescaling clocks and rulers at each space-time point. 

On the other hand, the process of theory building itself can be sensitive to the choice of the field variables. When constructing any theory, one relies on a set of assumptions and formulate criteria to represent these. If the criteria are more restrictive than the assumptions, the resulting theory will still be compatible with the assumptions but may not be a complete representation. 
One way to compensate this mismatch is to generalise the matter coupling to
\begin{equation}
\sqrt{-g}\,{\cal L}_{\rm vacuum} [g, \{\chi_a\}] + \sqrt{-\tilde{g}} \, {\cal L}_{\rm matter}[\tilde{g},\{\psi_b\}]\,,
\end{equation}
where $\{\chi_a\}$ represent the additional degrees of freedom that participate in the gravitational interactions, while $\{\psi_b\}$ are the matter degrees of freedom.\footnote{In our formulation, we assume a universal matter coupling. In the case where weak equivalence principle is broken, different matter sectors can flow on different geometries. Justification for a restricted version of this scenario is presented in Section \ref{sec:discussion}.}
Here matter follows the geodesics of the {\it Jordan frame} metric $\tilde{g}$, which can depend on all gravitational fields, i.e. metric $g$, fields $\{\chi_a\}$ and their derivatives. This generalisation goes back to Bekenstein's ``two geometries'' perspective.\footnote{In Bekenstein's nomenclature, $g$ is the gravitational metric while $\tilde{g}$ is the physical metric \cite{Bekenstein:1992pj}. Since matter follows the geodesics of $\tilde{g}$, the physical metric is uniquely defined. However, the interpretation of the gravitational metric is more ambiguous in modern modified gravity theories, where it is not always possible to define a field variable to reduce the vacuum action to general relativity and minimally coupled extra fields.
}
Although this approach was proposed as a ``method for constructing novel gravitational theories,'' \cite{Bekenstein:1992pj} its full strength started being acknowledged only recently, following the developments in scalar-tensor theories. 
In the presence of a single gravitational scalar degree of freedom $\phi$, the most general metric that depends on the original metric variable $g_{\mu\nu}$, the scalar field and its first derivatives is given via a general disformal relation \cite{Bekenstein:1992pj}
\begin{equation}
\tilde{g}_{\mu\nu} = C(\phi ,\partial_\alpha\phi\,\partial^\alpha\phi)\,g_{\mu\nu} + D(\phi,\partial_\alpha\phi\,\partial^\alpha\phi)\,\partial_\mu\phi\,\partial_\nu\phi\,,
\label{eq:generaldisformal}
\end{equation}
where the first term provides a conformal rescaling of the metric, while the second one, the {\it disformal} term, provides an anisotropic deformation that aligns with the field flow.
Provided that the transformation is invertible, different representations of a scalar-tensor theory are dynamically equivalent.\footnote{Although representations related by a change of variable are classically equivalent (see e.g. \cite{Flanagan:2004bz, Deruelle:2010ht, Domenech:2015hka}), subtle differences arise in their interpretations \cite{Sotiriou:2007zu}, while quantum anomalies may invalidate the physical equivalence \cite{Herrero-Valea:2016jzz}.
} Therefore, once the general form of the vacuum theory is obtained, it should not be sensitive to the field variable used for the matter coupling. 
However, as we discussed above, this is only true if the set of assumptions are represented accurately. 

This point becomes evident in the context of Horndeski's scalar-tensor theory \cite{Horndeski:1974wa, Deffayet:2011gz}, which relies on the assumption that the number of initial conditions necessary to evolve the system of dynamical equations of motion is simply $6$ for gravitational sector, i.e. only the scalar field and metric perturbations are dynamical. In the original construction, the theoretical formulation of this assumption includes the requirement that the equations of motion are at most of second order. This is based on Ostrogradski's result that for non-degenerate systems, 
more than two derivatives in the equations of motion introduce new degrees of freedom which lead to an instability \cite{Ostrogradsky:1850fid}. However, this representation, although it does not contradict with the assumption regarding the necessary initial data, is not exhaustive enough to cover all allowed interactions. 
One way to see this is to apply the general disformal transformation \eqref{eq:generaldisformal} to Horndeski action. 

Horndeski action is closed under special disformal transformations with $C=C(\phi), D=D(\phi)$ \cite{Bettoni:2013diz} but 
generalising the coefficients to those that depend on field derivatives leads to new terms, which are not present in the original formulation \cite{Zumalacarregui:2013pma}. With respect to the original variables, the general coupling manifestly keeps the equations of motion second order. However, in the Jordan frame, these interactions generate higher order equations of motion, seemingly hinting at an additional degree of freedom. This apparent inconsistency is resolved by the degeneracies, which reveal that Horndeski's selection criterion for initial data is stronger than necessary. Despite the existence of high derivatives in the equations of motion, the instability is absent thanks to a degenerate kinetic matrix in the Lagrangian, which produces a hidden constraint and removes the unwanted degree of freedom. Terms that extend Horndeski theory in this way were identified in the Beyond Horndeski theory \cite{Gleyzes:2014dya,Gleyzes:2014qga} and later, DHOST theories \cite{Langlois:2015cwa,BenAchour:2016fzp}. These scalar-tensor actions are now closed under transformation \eqref{eq:generaldisformal}, thus the choice of Jordan frame does not affect the generality of the theory (see a review \cite{Langlois:2018dxi} and references therein).

For modified gravity theories with multiple extra degrees of freedom, similar degenerate terms are more difficult to identify. In principle, constraint analysis techniques can be adopted to determine the fate of some given interaction terms, but they are not feasible to uncover all possible interactions exhaustively. Moreover, other extensions can be devised by coupling matter to a metric that cannot be transformed into Jordan frame via a simple field redefinition.

In this paper, we propose the use of a generalised matter coupling as a systematic, tractable and exhaustive theory building tool that preserves the compatibility between assumptions and their theoretical representations. As a first application, we consider the case of Lorentz invariant massive gravity theory with de Rham, Gabadadze, Tolley (dRGT) potential \cite{deRham:2010ik,deRham:2010kj}. In this theory, construction of exact cosmological solutions have proved to be  challenging, since either the expansion decouples from the matter sector \cite{DAmico:2011eto,Gumrukcuoglu:2011ew,Gumrukcuoglu:2011zh} or a non-linear ghost instability appears \cite{DeFelice:2012mx,DeFelice:2013awa}. This cosmological no-go result can be evaded by adding new degrees of freedom and/or relaxing the symmetries.\footnote{See \cite{deRham:2014zqa} and references therein.} Instead, in this paper we will explore extensions of the theory without changing its building blocks, by generalising the matter coupling. 
As opposed to the simple scalar-tensor theories, the stability of the construction is not guaranteed: the dRGT action is protected by several uniqueness theorems while also relying on a delicate constraint to reduce the number of propagating degrees of freedom down to five. In particular, a modified matter coupling is expected to remove the constraint. Our aim in this paper is to find new interaction terms that approximately preserve the dRGT constraint within the strong coupling scale. Such an example of an effective theory with a cut-off above the strong coupling scale was introduced in Ref.\cite{deRham:2014naa} by requiring that the quantum corrections do not detune the potential.
Our approach provides a new perspective that can identify generalisations of this example.

The paper is organised as follows. In the next section, we give a brief review of dRGT massive gravity focusing on the interaction scales and summarising the various uniqueness theorems. In Section \ref{sec:multigeometry}, we develop a generalisation of disformal transformations to the case of massive gravity. In Section \ref{sec:decouplinglimit}, we use the decoupling limit to identify and eliminate low energy interactions. We consider a simple example with constant coefficients in Section \ref{sec:composite}. We conclude with Section \ref{sec:discussion} where we discuss our results.
The paper is supplemented by four appendices where we summarise the technical steps.

\section{dRGT potentials: Scales, interactions and uniqueness theorems}
\label{sec:dRGT}
We start with a brief summary of massive gravity, focusing on the relevant scales that correspond to the interaction terms. For a complete and detailed review, see Ref.\cite{deRham:2014zqa}.

The mass term for a spin--2 field is written as an interaction between the metric tensor and a fixed reference metric, typically chosen as the Minkowski metric. Using the gravitational analogue of the St\"uckelberg trick, one can introduce four scalar fields $\phi^a$ to restore the diffeomorphism invariance. This promotes the reference metric to a space-time tensor given by
\begin{equation}
f_{\mu\nu} \equiv \eta_{ab}\partial_\mu\phi^a\,\partial_\nu\phi^b\,.
\end{equation}
In this formulation, $\eta_{ab}$ becomes the field space metric associated with a Poincar\'e symmetry. Mass terms can then be written in terms of scalar functions of the tensor $g^{-1}f$. A generic massive gravity theory has the following form
\begin{equation}
S = \frac{M_p^2}{2} \int d^4x \sqrt{-g}\left[ R[g] + m^2 {\cal F}(f^{-1}g)\right]\,,
\label{eq:general-massive-gravity}
\end{equation}
where we only consider non-derivative interactions between the two metrics. 
The graviton mass provides the kinetic term for the St\"uckelberg fields and there are thus four new dynamical degrees of freedom in addition to the two in general relativity. This is one more than the number of degrees of freedom of a spin--2 field required by Poincar\'e representations. The sixth degree of freedom, the Boulware-Deser mode, allows arbitrarily large negative energy \cite{Boulware:1972zf,Boulware:1973my}. This extra mode can be isolated as the longitudinal perturbation of the scalar fields
\begin{equation}
\phi^a  = x^a +\frac{\partial^a\pi}{M_pm^2}\,,
\end{equation}
where the background $\phi^a=x^a$ corresponds to the gauge where the $f$--metric coincides with Minkowski space-time, and we introduced the canonically normalised longitudinal perturbation. The additional derivative that accompanies $\pi$ in this decomposition allows its interpretation as the Boulware-Deser ghost, which manifests itself through an Ostrogradski instability of the helicity-0 mode. Around the Minkowski background $g=\eta$, the two--metric coupling is
\begin{equation}
g^{-1}f = \left(\mathbb{1}+\frac{\partial\partial\pi}{M_pm^2}\right)^2\,,
\end{equation}
where $\partial\partial\pi$ denotes the Hessian matrix of $\pi$. In general, an arbitrary function $\mathcal{F}$ in \eqref{eq:general-massive-gravity} would lead to interaction terms of the form 
\begin{equation}
\frac{1}{\left(M_p m^{\frac{2(n-1)}{n-2}}\right)^{n-2}}(\partial^2\pi)^n\,,
\end{equation}
which involve more than two derivatives of $\pi$. These dangerous interactions in the generic theory appear at the relatively low scale $\Lambda_5 \equiv (M_pm^4)^{1/5}$, which corresponds to a distance of $10^{11}{\rm km}$ for a present-day Hubble scale mass. 

Using the square-root of the tensor $g^{-1}f$ as a building block provides a more natural way to determine the conditions for the mass function, since this combination allows one to keep track of the dangerous terms themselves rather than their matrix square. The sixth mode can be removed by a new constraint \cite{Hassan:2011hr} a result of the dRGT potential \cite{deRham:2010ik,deRham:2010kj,Hassan:2011vm}
\begin{equation}
S = \frac{M_p^2}{2} \,\int d^4x \sqrt{-g}\left[ R[g] + m^2 \sum_{i=0}^{4}\,\beta_i \,e_i\left(\sqrt{g^{-1}f}\right)\right]\,,
\label{eq:dRGT-massive-gravity}
\end{equation}
where the mass terms consist of the elementary symmetric polynomials of $\sqrt{g^{-1}f}$ defined as
\begin{align}
e_0(x) &= 1\,,\nonumber\\
e_1(x) &= [x]\,,\nonumber\\
e_2(x) &= \frac{1}{2!}\,([x]^2-[x^2])\,,\nonumber\\
e_3(x) &= \frac{1}{3!}\,([x]^3-3[x][x^2]+2[x^3])\,,\nonumber\\
e_4(x) &= \frac{1}{4!}\,([x]^4-6[x]^2[x^2]+8[x][x^3]+3[x^2]^2-6[x^4])\,,
\label{eq:elementarypoly}
\end{align}
where square brackets denote the trace operation.
Although the action \eqref{eq:dRGT-massive-gravity} has six parameters, $\beta_0$ is the cosmological constant for the $g$ metric, $\beta_4$ simply generates a cosmological constant for the $f$--metric which does not affect the equations of motion \cite{Hassan:2011vm}, one combination corresponds to a tadpole term which is removed to allow Minkowski metric as a solution, and finally one parameter can be absorbed into $m^2$. As a result, the potential introduces three independent parameters, including $m$. Removing the ghost mode raises the strong coupling scale of the theory to $\Lambda_3\equiv (M_pm^2)^{1/3}$ which is about 8 orders of magnitude improvement compared to the generic massive theory.

The dRGT potential is the unique non-linear completion of the Pauli-Fierz mass term \cite{Comelli:2012vz}. For the kinetic part, a ghost-free kinetic term beyond the Einstein-Hilbert action has been discovered perturbatively \cite{Hinterbichler:2013eza}, although non-linear completion reintroduces the BD mode and leaves the Einstein-Hilbert term as the unique non-linear derivative term \cite{deRham:2013tfa} (see e.g. Refs.\cite{Kimura:2013ika, Tukhashvili:2017eqc} for other attempts, and Ref.\cite{Bonifacio:2018vzv} for an argument based on tree level scattering amplitudes). Finally, a matter field can minimally couple only to a single metric, whereas a more complicated coupling inevitably reintroduces the BD mode \cite{Yamashita:2014fga,deRham:2014naa,Heisenberg:2015iqa}. 

These uniqueness theorems for Lorentz invariant non-linear massive gravity crucially rely on the requirement that the constraint that removes the Boulware-Deser mode at all scales. One loophole exists: dRGT massive gravity is already an effective field theory valid up to a cutoff scale above $\Lambda_3$. By relaxing the condition to avoid this mode such that its mass is above $\Lambda_3$, one can obtain an extension of dRGT where the ghost is irrelevant in the decoupling limit. A specific example along this line was introduced in Ref.\cite{deRham:2014naa} by requiring that matter loops do not detune the dRGT potential at one loop order. This special coupling, which we will call ``composite coupling,'' generates the BD mode, but the mass is larger than $\Lambda_3$ in general and infinite around FLRW. 
\footnote{The mass of the ghost depends on the background configuration, and can become light around strong gravitational backgrounds. However, incorporating the non-linear effects keep their mass above $M_p$ \cite{deRham:2014naa}. Moreover, for the bimetric extension of the composite coupling, a tri-metric theory can provide a ghost-free completion \cite{Luben:2018kll}.} 
In the following, we will systematically make use of this loophole and look for generalisations of the composite coupling.

\section{Formulation of massive gravity with generalised matter coupling} 
\label{sec:multigeometry}
Our approach to obtain new interactions in a modified gravity theory is to first consider the vacuum theory with a given metric variable, then to minimally couple matter to a new geometry that is disformally related to the first one. This new geometry needs to be constructed using the non-metric degrees of freedom specific to the modified gravity theory in question. Provided that the coupling itself does not include higher derivatives of these fields, the matter coupling should preserve the number of degrees of freedom of the original theory.

In this section, we present the application of this approach to dRGT massive gravity theory by generalising the matter coupling to include direct couplings to the additional degrees of freedom. These degrees of freedom can be isolated in the St\"uckelberg picture as four scalar fields $\phi^a$ with an internal Poincar\'e symmetry. Thus to implement our approach, we first need to extend the disformal relation \eqref{eq:generaldisformal} to four scalar fields. We now re-formulate the theory to include an arbitrary matter coupling and derive the conditions for invertibility of the disformal relation.

\subsection{Formalism}
\label{sec:formalism}

We start by extending the disformal relations to four scalar fields in a straightforward way, while making use of the Poincar\'e symmetry in the field space. The latter simply means that the relation should not expose any free field indices. With this in mind, we propose the following generalisation:
\begin{equation}
\tilde{g}_{\mu\nu} = \bar C([\gamma^n]) g_{\mu\nu} + \bar D_{ab}([\gamma^n]) \nabla_\mu\phi^a\nabla_\nu\phi^b\,,
\label{eq:naive4disformal}
\end{equation}
where the functions $\bar C$ and $\bar D_{ab}$ depend on the traces of powers of space-time tensor $\gamma^\mu_{\;\;\nu} \equiv \eta_{ab}\nabla^\mu\phi^a\,\nabla_\nu\phi^b$. Since we have four scalar fields in four space-time dimensions, only the first four of these traces are independent, so the arguments of these functions are explicitly $\gamma^\mu_{\;\;\mu}$, $\gamma^\mu_{\;\;\nu}\gamma^\nu_{\;\;\mu}$, $\gamma^\mu_{\;\;\nu}\gamma^\nu_{\;\;\rho}\gamma^\rho_{\;\;\mu}$ and $\gamma^\mu_{\;\;\nu}\gamma^\nu_{\;\;\rho}\gamma^\rho_{\;\;\sigma}\gamma^\sigma_{\;\;\mu}$. 
Similarly, there are three independent ways of writing the function $\bar D_{ab}$: $\eta_{ab}$, $\eta_{ac}\eta_{bd}\nabla_\alpha\phi^c\nabla^\alpha\phi^d$,  $\eta_{ac}\eta_{bf}\eta_{de} \nabla_\alpha\phi^c\nabla^\alpha\phi^d\nabla_\beta\phi^e\nabla^\beta\phi^f$  so the disformal part can be equivalently written as 
\footnote{Other contractions with higher powers of $\nabla\phi$ do not produce any more independent terms. For instance, including the term 
$\eta_{ac}\eta_{bh}\eta_{de}\eta_{fg} \nabla_\alpha\phi^c\nabla^\alpha\phi^d\nabla_\beta\phi^e\nabla^\beta\phi^f\nabla_\gamma\phi^g\nabla^\gamma\phi^h$  in $\bar{D}_{ab}$ gives rise to $(g\gamma^4)_{\mu\nu}$ in \eqref{eq:naive4disformal}. However, according to the Cayley-Hamilton theorem, 
in 4 dimensions powers of $\gamma$ higher than 3 can be written as a power series with coefficients that depend on the characteristic polynomials \eqref{eq:elementarypoly} $e_n(\gamma)$ (or equivalently, on $[\gamma^n]$), i.e. 
\begin{equation}
\gamma^4 = e_1(\gamma) \gamma^3-e_2(\gamma)\gamma^2+e_3(\gamma)\gamma-e_4(\gamma) \mathbb{1}\,.
\end{equation}
This relation can be used to show that any other disformal construction will be one of the ones given in \eqref{eq:gtilde-generalised}.
}
\begin{equation}
\bar D_{ab}([\gamma^n]) \nabla_{\mu}\phi^a\nabla_{\nu}\phi^b= \bar D([\gamma^n])(g\gamma)_{\mu\nu} + \bar E([\gamma^n])(g\gamma^2)_{\mu\nu}+ \bar F([\gamma^n])(g\gamma^3)_{\mu\nu}\,.
\label{eq:gtilde-generalised}
\end{equation}
Thus the most generic transformation is 
\footnote{In order to preserve the symmetries of the dRGT potential, we imposed invariance under translations $\phi^a\to\phi^a+c^a$. If one relaxes this assumption in the fashion of Ref.\cite{deRham:2014gla}, the functions $C-F$ would also depend on combinations that include the fields themselves, e.g. $\eta_{ab}\phi^a\phi^b$, or mixed traces such as $\eta_{ad}\phi^a\phi^b\eta_{bc}\nabla_\mu \phi^c\nabla^\mu \phi^d$. Moreover, the independent disformal terms in $D_{ab}$ would acquire new contributions, e.g. $\eta_{ac}\eta_{bd}\phi^c\phi^d$.
}
\begin{equation}
\tilde{g}_{\mu\nu} = \bar C([\gamma^n]) g_{\mu\nu} + \bar D([\gamma^n]) f_{\mu\nu}+ \bar E([\gamma^n]) (g \gamma^2)_{\mu\nu}+ \bar F([\gamma^n]) (g \gamma^3)_{\mu\nu}\,,
\label{eq:nosquareroot-disformal}
\end{equation}
where $f_{\mu\nu}\equiv \eta_{ab} \partial_\mu \phi^a\partial_\nu\phi^b$ and $\gamma \equiv g^{-1} f$.
Although the above relation is the most general four-field extension of \eqref{eq:generaldisformal} that involves first derivatives and an internal Poincar\'e symmetry, it is not unique. In the context of dRGT massive gravity, it is more convenient to adopt an alternative formulation that replaces all occurrences of $\gamma$ with $\sqrt{\gamma}$ 
 as follows:\footnote{The relation \eqref{eq:mostgeneraltilde} can also be obtained by starting with a Finslerian geometry and requiring it to reduce to a Riemannian one \`a la Bekenstein (see Appendix \ref{app:proofofdisformal}). The equivalence between \eqref{eq:nosquareroot-disformal} and  \eqref{eq:mostgeneraltilde} is shown in Appendix \ref{app:squarerootformalism}. }
\begin{equation}
\tilde{g}_{\mu\nu} = C([\sqrt{\gamma}^n]) g_{\mu\nu} + D([\sqrt{\gamma}^n]) (g\sqrt{\gamma})_{\mu\nu}+ E([\sqrt{\gamma}^n]) f_{\mu\nu}+ F([\sqrt{\gamma}^n]) (f\sqrt{\gamma})_{\mu\nu}\,.
\label{eq:mostgeneraltilde}
\end{equation}
This formulation will be adopted in the rest of the text, on the basis that it trivially contains the composite matter coupling introduced in Ref.\cite{deRham:2014naa}.

The action for this construction is
\begin{equation}
S = \frac{M_p}{2}\,\int d^4x \sqrt{-\tilde{g}} \,\left[R[\tilde{g}] + m^2\sum_{i=0}^3 \beta_i e_i(\sqrt{\tilde{\gamma}})\right]+ \int\,\sqrt{-g}\,\mathcal{L}_{matter}\,,
\label{eq:action}
\end{equation}
where $\tilde{\gamma}= \tilde{g}^{-1}f$.

The uniqueness theorems imply that a generic coupling to $g=g(\tilde{g},f)$ would reintroduce the Boulware-Deser instability. In the Jordan frame this corresponds to new interactions, such as derivative interactions  that depend on the difference of connections for $g$ and $f$ metrics, $\Gamma(g) - \Gamma(f)$ (see Appendix \ref{app:jordan}).

\subsection{Invertibility}
\label{sec:invertibility}
The invertibility of the disformal relation \eqref{eq:mostgeneraltilde} is crucial in determining whether this is really a field redefinition. A transformation that is not invertible simply means that the new variable does not contain sufficient information to reconstruct the old one, and corresponds to an implicit choice of a preferred frame.

If the Jacobian of the transformation 
\begin{equation}
J^{\alpha\beta}_{\;\;\;\;\mu\nu} \equiv\frac{\partial\tilde g_{\mu\nu}}{\partial g_{\alpha\beta}}\,,
\label{eq:jacobian-formal}
\end{equation}
has no zero eigenvalues, the transformation is conjectured to be invertible \cite{Zumalacarregui:2013pma}. For the case at hand, determining the Jacobian involves taking derivatives of the square-root tensor $X^{\mu}_{\;\;\nu}$. Starting from the definition of $X$,
\begin{equation}
g^{-1}f = X^2\,,
\end{equation}
then differentiating both sides, we have
\begin{equation}
X\,\delta X+\delta X X =-g^{-1} \delta g\,X^2\,,
\end{equation}
which is a matrix equation for $\delta X$ of the Sylvester type, the analytic solutions of which are known \cite{HU2006859}. Using the form of the solutions given in Ref.~\cite{Bernard:2015mkk}, we find
\begin{equation}
\delta X = -\frac{1}{2}(e_1X^2+e_3 \mathbb{1})^{-1}\sum_{k=1}^{4}\sum_{m=0}^{k-1}(-1)^me_{4-k}X^{k-m-2}g^{-1} \delta g\,X^{m+2}\,,
\end{equation}
where $e_n$ are the characteristic polynomials corresponding to $X$. Note that the solution is unique provided that the matrix $(e_1X^2+e_3 \mathbb{1})$ is invertible \cite{Bernard:2015mkk}. Using this solution, we can write down the Jacobian as
\begin{equation}
J^{\alpha\beta}_{\;\;\;\;\mu\nu} =\mathfrak{C}^{\alpha\beta}g_{\mu\nu}+\mathfrak{D}^{\alpha\beta}(g X)_{\mu\nu} + \mathfrak{E}^{\alpha\beta}f_{\mu\nu}+\mathfrak{F}^{\alpha\beta}(f X)_{\mu\nu}+ C\, \delta^{(\alpha}_\mu \delta^{\beta)}_\nu+D\,\delta^{(\alpha}_\mu X^{\beta)}_{\;\;\nu}+(D\,Q_{\mu\rho}+F\,\bar{Q}_{\mu\rho})\Delta^{\rho\alpha\beta}_{\;\;\;\;\;\;\nu}\,,
\label{eq:jacobian-exact}
\end{equation}
where brackets around indices denote normalised symmetrisation, and we defined
\begin{align}
Q_{\mu\nu}\equiv &\frac{1}{e_1^2e_4+e_3^2-e_1e_2e_3} \,g_{\mu\alpha} \left[(e_3-e_1e_2)\delta^{\alpha}_\nu+e_1^2 X^\alpha_{\;\;\nu}-e_1 (X^2)^{\alpha}_{\;\;\nu}\right]\,,\nonumber\\
\bar{Q}_{\mu\nu}\equiv &\frac{1}{e_1^2e_4+e_3^2-e_1e_2e_3} \,f_{\mu\alpha} \left[(e_3-e_1e_2)\delta^{\alpha}_\nu+e_1^2 X^\alpha_{\;\;\nu}-e_1 (X^2)^{\alpha}_{\;\;\nu}\right]\,,\nonumber\\
\Delta^{\rho\alpha\beta}_{\;\;\;\;\;\;\nu} \equiv& -\frac{1}{2}\,\sum_{k=1}^4\sum_{m=0}^{k-1}(-1)^me_{4-k}(X^{k-m-2})^\sigma_{\;\;\tau} g^{\tau(\alpha}(X^{m+2})^{\beta)}_{\;\;\nu}\,.
\end{align}
In Eq.\eqref{eq:jacobian-eigenproblem} the Fraktur letters denote derivatives of the coefficients with respect to the metric $g_{\mu\nu}$. For instance, for the derivative of $C$, one has:
\begin{equation}
\mathfrak{C}^{\alpha\beta}\equiv \frac{\partial C([X],[X^2],[X^3],[X^4])}{\partial g_{\alpha\beta}}
=-\frac{1}{2}\sum_{m=1}^4\,m\,C_m g^{\rho(\alpha}(X^m)^{\beta)}_{\;\;\rho}\,
\end{equation}
where $C_m$ is the derivative of $C$ with respect to its $m$--th argument 
\begin{equation}
C_m \equiv \frac{\partial C([X],[X^2],[X^3],[X^4])}{\partial [X^m]}]\,.
\end{equation}
Similar definitions apply to the remaining coefficients in \eqref{eq:mostgeneraltilde}.
 
 Given the exact expression for the Jacobian \eqref{eq:jacobian-exact}, one can obtain the invertibility conditions by solving the following eigenvalue problem \cite{Zumalacarregui:2013pma}
\begin{equation}
\left(J^{\alpha\beta}_{\;\;\;\;\mu\nu} -\lambda \,\delta^{(\alpha}_\mu\delta^{\beta)}_\nu\right)\xi_{\alpha\beta}=0\,.
\label{eq:jacobian-eigenproblem}
\end{equation}
Considering the symmetries of the Jacobian, there are $10$ eigenvalues $\lambda_n$ which should satisfy
\begin{equation}
\prod_{n=1}^{10}\lambda_n \neq0\,,
\label{eq:invertibility-generic}
\end{equation}
for the transformation to be invertible. 

\section{Stability conditions in the decoupling limit}
\label{sec:decouplinglimit}
In the previous Section, we formulated the dRGT theory with a generic matter coupling. Based on the arguments in Section \ref{sec:dRGT}, this coupling reintroduces the Boulware-Deser mode. Our goal in this Section is to push this mode beyond the strong coupling scale $\Lambda_3$. To accomplish this, we will tune the unknown functions in order to remove all dangerous interactions in the decoupling limit. This requires a perturbative treatment. 

Moreover, our aim is to obtain conditions that are independent on the details of the matter sector. Therefore our analysis will be carried out in the Jordan frame, where we do not need to specify individual matter fields.

\subsection{Decomposing the fields}
\label{sec:Perturbations}
We now consider scalar perturbations around flat space-time. We define the Minkowski vacuum in the dRGT frame, i.e.
\begin{equation}
\tilde{g}_{\mu\nu}=\eta_{\mu\nu}+\tilde{h}_{\mu\nu}\,.
\end{equation}
For the St\"uckelberg fields, we will only concentrate in the longitudinal perturbation, i.e. $\phi^a = x^a + \partial^a\pi$, which can be used to identify the Boulware-Deser ghost through an Ostrogradski instability. With this decomposition, the $f_{\mu\nu}$ tensor becomes
\begin{equation}
f_{\mu\nu} = \eta_{\mu\nu} +\Pi_{\mu\nu}+\Pi_{\nu\mu} +\Pi_{\mu}^{\;\;\alpha}\Pi_{\nu\alpha}\,,
\label{eq:fperturbations}
\end{equation}
where $\Pi_{\mu\nu} \equiv \partial_\mu\partial_\nu\pi$. Once the vacuum is selected, the background for the Jordan frame metric can only be conformally Minkowski in general, which we decompose as
\begin{equation}
g_{\mu\nu} = \Omega_0(\eta_{\mu\nu}+h_{\mu\nu})\,.
\label{eq:gperturbations}
\end{equation}
At quartic order in perturbations, only a finite number of combinations of functions $C$, $D$, $E$, $F$ and their derivatives are relevant. The explicit form of these combinations are given in Appendix \ref{app:coefficients}.

\subsection{Invertibility revisited: perturbative case}
The discussion in Section \ref{sec:invertibility} reveals that the Jacobian of transformation \eqref{eq:mostgeneraltilde} has ten eigenvalues, given by the roots of Eq.\eqref{eq:jacobian-eigenproblem}. Solving this equation generically is not straightforward. However, the invertibility condition does not have to be satisfied for arbitrary configurations, but rather should be determined depending on the context. For the present perturbative discussion, showing invertibility for the background is sufficient. The Minkowski background provides a dramatic simplification since $g_{\mu\nu}\propto f_{\mu\nu}$. As a result, the Jacobian has only two independent eigenvalues. For the background discussed in Sec.\ref{sec:Perturbations}, we have $g_{\mu\nu}=\Omega_0\eta_{\mu\nu}$ and $f_{\mu\nu} = \eta_{\mu\nu}$, which lead to 
\begin{align}
X^{\mu}_{\;\;\nu}\Big\vert_{f=g/\Omega_0=\eta}=&\Omega_0^{-1/2}\,\delta^{\mu}_\nu\,,\nonumber\\
Q_{\mu\nu}\Big\vert_{f=g/\Omega_0=\eta} =& \frac{\Omega_0^{5/2}}{8}\,\eta_{\mu\nu}\,,\nonumber\\
\bar{Q}_{\mu\nu}\Big\vert_{f=g/\Omega_0=\eta} =& \frac{\Omega_0^{3/2}}{8}\,\eta_{\mu\nu}\,,\nonumber\\
\Delta^{\rho\alpha\beta}_{\;\;\;\;\;\;\nu}\Big\vert_{f=g/\Omega_0=\eta}=& -\frac{4}{\Omega_0^{3}}\,\eta^{\rho(\alpha}\delta^{\beta)}_{\nu}\,.
\end{align}
Using these, we can rewrite the Jacobian \eqref{eq:jacobian-exact} as
\begin{equation}
J^{\alpha\beta}_{\;\;\;\;\mu\nu}\Big\vert_{f=g/\Omega_0=\eta} = \frac{\mathcal{B}_1-4\,\mathcal{B}_2}{3\,\Omega_0}\eta_{\mu\nu}-\frac{\mathcal{A}_1}{\Omega_0}\,\delta^{(\alpha}_\mu\delta^{\beta)}_\nu\,,
\end{equation}
where $\mathcal{A}_1$, $\mathcal{B}_1$ and $\mathcal{B}_2$ depend on the values of the coefficients and their first derivatives, and they are defined in Appendix \ref{app:coefficients}. For this background, the eigenvalue problem \eqref{eq:jacobian-eigenproblem} can be solved by 
\begin{equation}
\lambda\Big\vert_{f=g/\Omega_0=\eta}= \left\{
\begin{array}{lll}
\dfrac{-3\,\mathcal{A}_1+4\,\mathcal{B}_1-16\,\mathcal{B}_2}{3\,\Omega_0}& \,,\;\xi_{\mu\nu}\propto \eta_{\mu\nu}& \,,\;(1 {\rm ~eigenvalue})\\
-\dfrac{\mathcal{A}_1}{\Omega_0}& \,,\;\xi_{\mu\nu}\eta^{\mu\nu}=0& \,,\;(9 {\rm ~eigenvalues})
\end{array}\right.
\label{eq:jacobian-eigenvalues-minkowski}
\end{equation}
Thus the invertibility condition \eqref{eq:invertibility-generic} for the Minkowski background simply becomes
\begin{equation}
\frac{\mathcal{A}_1^9(3\,\mathcal{A}_1-4\,\mathcal{B}_1+16\,\mathcal{B}_2)}{3\,\Omega_0^{10}}\neq0\,.
\label{eq:invertibility-minkowski}
\end{equation}
\subsection{Background consistency and the free field action}
The consistency of the relation \eqref{eq:mostgeneraltilde} imposes two conditions at the background level. The first is the consistency of the solutions which can be summarised as 
\begin{equation}
\mathcal{A}_2=1\,,
\label{eq:background-consistency}
\end{equation}
which is an equation that determines the value of $\Omega_0$. The second condition is the invertibility of the transformation which imposes the inequality \eqref{eq:invertibility-minkowski},
\begin{equation}
\mathcal{A}_1(3\,\mathcal{A}_1-4\,\mathcal{B}_1+16\,\mathcal{B}_2)\neq0\,.
\label{eq:invertibility-minkowski-re}
\end{equation}

We now move on to the action. At linear order, we have
\begin{equation}
\delta^{(1)}\mathcal{L} = -\frac{m^2M_p^2\,h}{12}(\beta_0+3\,\beta_1+3\,\beta_2+\beta_3)(3\,\mathcal{A}_1-4\mathcal{B}_1+16\,\mathcal{B}_2)\,.
\label{eq:linearact}
\end{equation}
For a consistent Minkowksi background, the variation of the action with respect to the metric perturbations $h_{\mu\nu}$ should vanish. The last factor in the linear term \eqref{eq:linearact} coincides with part of the invertibility condition \eqref{eq:invertibility-minkowski-re}
so it cannot be zero.
Therefore, the existence of the background solution requires the parameters to satisfy
\begin{equation}
(\beta_0+3\,\beta_1+3\,\beta_2+\beta_3)=0\,,
\label{eq:tadpole}
\end{equation}
which simply removes the tadpole term.\footnote{In the notation of Ref.\cite{deRham:2010kj}, this condition is simply $\alpha_1=0$. Note that the correspondence between the mass terms $\beta_n e_n(\sqrt{\gamma})$ and $\alpha_n e_n(\sqrt{\gamma}-\mathbb{1})$ is $\beta_0=-4\,\alpha_1+6-4\,\alpha_3+\alpha_4$, $\beta_1 = \alpha_1-3+3\,\alpha_3-\alpha_4$, $\beta_2 = 1-2\,\alpha_3+\alpha_4$, $\beta_3 = \alpha_3-\alpha_4$.}

The free field action, which is quadratic in perturbations, is obtained as
\begin{align}
\delta^{(2)}{\cal L} =&\frac{\mathcal{A}_1^2 M_p^2}{8}\Big\{
\partial^\mu h\partial_\mu h-\partial^\mu h^{\alpha\beta}\partial_\mu h_{\alpha\beta}-2\,\partial^\mu h\partial^\nu h_{\mu\nu}+2\,\partial^\mu h^{\nu\rho}\partial_\rho h_{\mu\nu} 
\nonumber\\
&\qquad\qquad
+ \frac{m^2(\beta_0+2\,\beta_1+\beta_2)}{2}\left[(h^2-h_{\mu\nu}h^{\mu\nu})+4\,(h_{\mu\nu}\partial^\mu\partial^\nu \pi - h\,\Box \pi)\right]
\Big\}
\nonumber\\
&+\frac{(\mathcal{B}_1-4\,\mathcal{B}_2)\,M_p^2}{12}\Big\{
(\mathcal{B}_1-4\,\mathcal{B}_2)\,\partial_\mu(h-2\,\Box\pi)\,\partial^\mu(h-2\,\Box\pi)-2\,\mathcal{A}_1(\partial^\mu h-\partial_\nu h^{\mu\nu})\,\partial_\mu (h-2\,\Box\pi)
\nonumber\\
&\qquad\qquad\qquad\qquad
+\frac{m^2(\beta_0+2\,\beta_1+\beta_2)(-3\,\mathcal{A}_1+2\,\mathcal{B}_1-8\,\mathcal{B}_2)}{2}(h-2\,\Box\pi)^2
\Big\}
\,.
\label{eq:generalquadratic}
\end{align}
We note that the last two lines contain four and six derivative terms of $\pi$. These can be removed by a field redefinition
\begin{equation}
h_{\mu\nu}\to h_{\mu\nu} -\frac{2(\mathcal{B}_1-4\,\mathcal{B}_2)}{3\,\mathcal{A}_1-4\,\mathcal{B}_1+16\,\mathcal{B}_2}\eta_{\mu\nu}\,\Box\pi\,.
\end{equation}
However, in the presence of matter, this shift replaces the high derivative to the matter coupling in the form of $\Box\pi \,T$ \cite{Naruko:2018akp}. To avoid instabilities due to this coupling, we choose
\begin{equation}
\mathcal{B}_2 = \frac{\mathcal{B}_1}{4}\,,
\label{eq:noquadraticghost}
\end{equation}
which effectively picks out the Fierz-Pauli action as the free theory.
We will impose this condition from here on. 

We also note that the mass in the free field action \eqref{eq:generalquadratic} always enters with the combination $m^2(\beta_0+2\,\beta_1+\beta_2)$. Therefore, without loss of generality, we can set
\begin{equation}
\beta_0+2\,\beta_1+\beta_2=1\,,
\end{equation}
which is equivalent to absorbing this term into the definition of $m$.\footnote{In the equivalent $e_n(\sqrt{\gamma}-\mathbb{1})$ formulation, this corresponds to fixing the coefficient of the $e_2(\sqrt{\gamma}-\mathbb{1})$ term to unity.}
Notice that \eqref{eq:noquadraticghost} also reduces the invertibility condition \eqref{eq:invertibility-minkowski} simply to $\mathcal{A}_1\neq0$. We normalise the two fields 
and perform a conformal transformation in the $h$ perturbations, via
\begin{equation}
h= -\frac{1}{\mathcal{A}_1 M_p}\left(h_c + \eta\,\pi_c\right)\,,\qquad
\pi = \frac{2}{\mathcal{A}_1 M_pm^2}\,\pi_c\,.
\end{equation}
This transformation decouples the kinetic terms of the metric perturbations from the St\"uckelberg scalar at the level of the quadratic action
\begin{align}
\delta^{(2)}{\cal L} =&\frac{1}{8}\Big\{
\partial^\mu  h\partial_\mu  h-\partial^\mu  h^{\alpha\beta}\partial_\mu  h_{\alpha\beta}-2\,\partial^\mu  h\partial^\nu  h_{\mu\nu}+2\,\partial^\mu  h^{\nu\rho}\partial_\rho  h_{\mu\nu} -6\,\partial_\mu\pi\,\partial^\mu\pi
\nonumber\\
&\qquad
- \frac{m^2}{2}\left[ h_{\mu\nu} h^{\mu\nu}- h^2+6\,\pi\,(h+2\,\pi)\right] \Big\}
\,,
\label{eq:quadratic}
\end{align}
where we suppressed the subscript ``$c$'' for the sake of clarity.

\subsection{Decoupling limit and interaction terms}

We can now extract the information about nonlinear interactions by going to the decoupling limit
\begin{equation}
m\to0\,,\;M_p\to\infty\,,\;\Lambda_3 \to {\rm finite}\,,
\end{equation}
where the quadratic action \eqref{eq:quadratic} simply becomes
\begin{equation}
\delta^{(2)}{\cal L}_{\rm D.L.} =\frac{1}{8}\left[
\partial^\mu h\partial_\mu h-\partial^\mu h^{\alpha\beta}\partial_\mu h_{\alpha\beta}-2\,\partial^\mu h\partial^\nu h_{\mu\nu}+6\,\partial^\mu h^{\nu\rho}\partial_\rho h_{\mu\nu} -4\,\partial^\mu h_{\mu\nu} \partial_\rho h^{\rho\nu}-6 \,\partial_\mu\pi\,\partial^\mu\pi\right]\,.
\end{equation}

We now determine at which order the dangerous high derivative terms will appear. After the transformation \eqref{eq:mostgeneraltilde}, the Einstein-Hilbert term contains generically terms with high derivatives schematically of the form $\partial^{2\,(n-a+1)}h^a\,\pi^{n-a}$, which are suppressed by $\Lambda_K ^{3n-2(a+1)}$, where
\begin{equation}
\Lambda_K \equiv (M_pm^{K-1})^{1/K}\,,\qquad {\rm with~~}K \equiv 3+\frac{2(2-a)}{n-2}\,.
\end{equation}
In order to avoid generating the ghost mode, we need to make sure that there are no high derivative interactions for $K>3$. Thus, any vertex which contains more than two $h_{\mu\nu}$ is beyond the reach of the effective theory. To be precise, at cubic order we expect interactions suppressed by $(\Lambda_{7-2a})^{7-2a}$, while at quartic order, interactions are suppressed by $(\Lambda_{5-a})^{2(5-a)}$ and so on. 

As for the mass term, the interactions are schematically of the form $\partial^{2(n-a)} h^a \pi^{n-a}$ and are suppressed by $\Lambda_P^{3n-2a-4}$, where
\begin{equation}
\Lambda_P \equiv (M_pm^{P-1})^{1/P}\,,\qquad {\rm with~~}P \equiv 3+\frac{2(1-a)}{n-2}\,.
\end{equation}
To avoid potential instabilities within the regime of validity of the effective field theory, we need to tune away all interactions below $\Lambda_3$ with $a<1$. In other words, we need to make sure that the vertices that contain at most one copy of $h_{\mu\nu}$ do not contribute to the dynamics. 

Unfortunately, the dangerous interactions that arise from the kinetic and potential terms appear at all orders in perturbation theory, thus the tuning needs to be done indefinitely. Below we demonstrate the procedure up to quartic interactions which, it turns out, is sufficient to reach a non-trivial conclusion regarding the form of the functions $C$, $D$, $E$ and $F$ in \eqref{eq:mostgeneraltilde}.

We start by discussing the cubic terms. The action cubic in perturbations is formally
\begin{align}
\delta^{(3)}{\cal L} = \frac{1}{\Lambda_5^5}\,\mathcal{L}^{(3)}_{\Lambda5} + \frac{1}{\Lambda_3^3}\,\mathcal{L}^{(3)}_{\Lambda3} + \mathcal{O}\left(\frac{m^{2/3}}{M_p^{2/3}}\right)
\end{align}
We see that the leading order in the decoupling limit is $\Lambda_5$, given by
\begin{align}
\mathcal{L}^{(3)}_{\Lambda5} =& -(\mathcal{Q}_1+\mathcal{Q}_3)h\,\partial_\mu\Pi^{\mu\nu}\partial_\nu\,\Pi + (\mathcal{Q}_1-\mathcal{Q}_4)h\,\partial_\mu\Pi_{\nu\alpha}\partial^\mu\Pi^{\nu\alpha}+2\,\mathcal{Q}_1 h^{\mu\nu}\partial_\nu \Pi_{\mu\alpha}\partial^\alpha \Pi
\nonumber\\
&+(-2\,\mathcal{Q}_1+\mathcal{Q}_4)h^{\mu\nu}\partial_\mu\Pi_{\alpha\beta}\partial_\nu\Pi^{\alpha\beta}+\mathcal{Q}_3 h^{\mu\nu}\partial_\mu\Pi\partial_\nu\Pi + (\mathcal{Q}_2-\mathcal{Q}_3)h\,\Pi \,\Box\Pi-(\mathcal{Q}_2+\mathcal{Q}_4)h\,\Pi^{\mu\nu}\partial_\mu\partial_\nu\Pi 
\nonumber\\
&-\mathcal{Q}_2h^{\mu\nu}\Pi_{\mu\nu}\Box\Pi+2\,\mathcal{Q}_2h^{\mu\nu}\Pi_{\mu\alpha}\partial_\nu\partial^\alpha\Pi-(\mathcal{Q}_2-\mathcal{Q}_3)h^{\mu\nu}\Pi\,\partial_\mu\partial_\nu\Pi +\mathcal{Q}_4h^{\mu\nu}\Pi^{\alpha\beta}\partial_\mu\partial_\nu\Pi_{\alpha\beta}\,,
\end{align}
where $\Pi_{\mu\nu}\equiv \partial_\mu\partial_\nu \pi$ and we defined
\begin{equation}
\mathcal{Q}_1 \equiv\frac{\mathcal{A}_1^2-3\,\mathcal{A}_1-4\,\mathcal{A}_3}{\mathcal{A}_1^2}
\,,\qquad
\mathcal{Q}_2\equiv \frac{4\,\mathcal{B}_3}{\mathcal{A}_1^2}
\,,\qquad
\mathcal{Q}_3\equiv \frac{8\,\mathcal{C}_1}{\mathcal{A}_1^2}
\,,\qquad
\mathcal{Q}_4\equiv \frac{4\,\mathcal{B}_1}{\mathcal{A}_1^2}\,.
\end{equation}
The terms suppressed by $\Lambda_5$ scale cannot be removed by adding boundary terms and they all contain six derivatives. The four linearly independent  coefficients $\mathcal{Q}_n$ vanish if
\begin{equation}
\mathcal{A}_3=\frac{\mathcal{A}_1(\mathcal{A}_1-3)}{4}\,,\;
\mathcal{B}_1=\mathcal{B}_3=\mathcal{C}_1=0\,.
\label{eq:noLambda5}
\end{equation}
With this choice, Eq.\eqref{eq:noquadraticghost} now implies $\mathcal{B}_2=0$. From here on, we adopt the conditions \eqref{eq:noLambda5}.

Finally, we calculate the $\Lambda_3$ terms. After adding appropriate boundary conditions, the $\Lambda_3$ terms in the cubic action can be reduced to six types of terms:  $\pi\, \partial\partial\pi\, \partial\partial\pi$, $h\, \partial\partial\pi\, \partial\partial\pi$, $\pi\, \partial\partial h\, \partial\partial\pi$, $h\, \partial\partial h\, \partial\partial\pi$, $\partial\partial h\, \partial\pi\, \partial\pi$, $\partial\partial \pi\, \partial h\, \partial h$, given by
\begin{align}
\mathcal{L}^{(3)}_{\Lambda3} =& \mathcal{P}_1 \,\left(\pi+\frac{h}{2}\right)\left[(\Box\pi)^2-\partial_\mu\partial_\nu\pi\, \partial^\mu\partial^\nu\pi\right] +(\mathcal{P}_1-2\,\mathcal{P}_2)h^{\mu\nu}\left[\partial_\mu\partial_\rho\pi \partial_\nu\partial^\rho \pi-\partial_\mu\partial_\nu\pi\,\Box\pi\right]
\nonumber\\
&-\mathcal{P}_2\,\pi\,\left[ \Box h_{\mu\nu}\partial^\mu\partial^\nu\pi - \partial_\mu\partial_\nu h^{\mu\nu}\,\Box\pi\right]
+2\,\mathcal{P}_2\,\left[\Box h\,\partial_\mu\pi\,\partial^\mu\pi - \partial_\mu\partial_\nu h^{\nu\rho}\,\partial^\mu\pi\partial_\rho\pi\right]
\nonumber\\
&
+2\,\mathcal{P}_2\left[\Box\pi\,\partial_\mu\partial^\nu h^{\mu\rho}h_{\nu\rho}-\partial_\mu\partial^\nu\pi\,\Box h^{\mu\rho}h_{\rho\nu} +\partial_\mu\partial_\nu\pi\,\partial_\rho\partial_\sigma h^{\mu\rho}\,h^{\sigma\nu}-\partial_\mu\partial_\nu \pi\,\partial_\rho\partial_\sigma h^{\rho\sigma} h^{\mu\nu}\right]
\nonumber\\
&+\mathcal{P}_2\,\left[-2\,\Box\pi\partial_\mu h^{\mu\nu} \partial_\nu h+\Box \pi \partial_\mu h \partial^\mu h+4\,\Box\pi \,\partial^\mu h_{\mu\nu}\partial_\rho h^{\rho\nu}- \Box\pi\, \partial_\mu h_{\nu\rho}\partial^\mu h^{\nu\rho}+2\,\partial_\mu\partial_\nu \pi\, \partial^\mu h^{\nu\rho}\partial_\rho h\right.
\nonumber\\
&
\qquad\quad
-6\,\partial_\mu\partial_\nu\pi\, \partial^\mu h^{\nu\sigma}\partial^\rho h_{\rho\sigma}
-2\,\partial_\mu\partial_\nu\pi\,\partial^\mu h \,\partial^\nu h
+2\, \partial_\mu\partial_\nu\pi \,\partial^\mu h_{\rho\sigma}\,\partial^\nu h^{\rho\sigma}
+4\,\partial_\mu\partial_\nu \pi\,\partial^\mu h\,\partial_\rho h^{\nu\rho}
\nonumber\\
&
\qquad\quad
\left.
-2\,\partial_\mu\partial_\nu\pi\,\partial^\rho h^{\mu\nu}\partial_\rho h
\right]\,,
\label{eq:cubic}
\end{align}
where the two linearly independent coefficients are defined as
\begin{equation}
\mathcal{P}_1\equiv\frac{1-\mathcal{A}_1(\beta_1+\beta_2)}{\mathcal{A}_1}\,,
\qquad
\mathcal{P}_2 \equiv \frac{\mathcal{A}_1+1}{4\,\mathcal{A}_1}\,. 
\end{equation}
In the dRGT limit, i.e. $\mathcal{A}_1=-1$, one has $\mathcal{P}_1=1-\alpha_3$ and $\mathcal{P}_2=0$. At first sight, the interaction terms with coefficients $\propto\mathcal{P}_2$ seem to give rise to high derivative equations of motion. In particular, the $hh\pi$ interactions that stem from the Einstein-Hilbert term contain four derivatives. However, these terms can be removed by a non-linear local transformation. Shifting the metric perturbations via 
\begin{equation}
h_{\mu\nu}\to h_{\mu\nu} -\frac{2\,\mathcal{P}_1}{\Lambda_3^3}\,\partial_\mu\pi\partial_\nu\pi+\frac{4\,\mathcal{P}_2}{\Lambda_3^3} \eta_{\mu\nu}\partial_\alpha\pi\,\partial^\alpha\pi +\frac{8\,\mathcal{P}_2}{\Lambda_3^3}\,\partial^\alpha\pi\left(\partial_\alpha h_{\mu\nu}-\partial_{(\mu}h_{\nu)\alpha}\right)\,,
\label{eq:nonlinearshift}
\end{equation}
we find that the $\Lambda_3$ suppressed term in the cubic action reduces to 
\begin{equation}
\mathcal{L}^{(3)}_{\Lambda3} =\mathcal{P}_1\,\pi \left[(\Box\pi)^2-\Pi_{\mu\nu}\Pi^{\mu\nu}\right]\,,
\end{equation}
which is simply the Galileon type self interaction. In vacuum, this shows the equivalence of the action \eqref{eq:action} to the standard dRGT theory. In the presence of matter, since the
shift \eqref{eq:nonlinearshift} contains up to first derivatives only, no instability from cubic order interactions arises.

Using the conditions for the absence of $\Lambda_5$ cubic interactions \eqref{eq:noLambda5}, we now move on to the interactions of quartic order in perturbations. There are two cases. From the mass term, we expect $\partial^8\pi^4$ term at $\Lambda_4$, while the kinetic part should bring $\partial^{10}\pi^4$ at $\Lambda_5$ and $\partial^8h\pi^3$ at $\Lambda_4$. We find that both $\pi^4$ terms are boundary terms and only the $\partial^8h\pi^3$ interaction remains. Formally, we have 
\begin{align}
\delta^{(4)}{\cal L} = \frac{1}{\Lambda_4^8}\,\mathcal{L}^{(4)}_{\Lambda4} + \frac{1}{\Lambda_3^6}\,\mathcal{L}^{(4)}_{\Lambda3} + \mathcal{O}\left(\frac{m^{2/3}}{M_p^{2/3}}\right)\,.
\end{align}
Up to boundary terms, we can compute the $\Lambda_4$ suppressed terms as
\begin{align}
\mathcal{L}^{(4)}_{\Lambda4} =&16\,\left[(\mathcal{R}_3-\mathcal{R}_7)\Pi^2 -(\mathcal{R}_2+2\,\mathcal{R}_4)\Pi_{\mu\nu}\Pi^{\mu\nu}\right]\left(h\,\Box\Pi - h_{\rho\sigma}\Box \Pi^{\rho\sigma}\right)
+2\,(\mathcal{R}_1+8\,\mathcal{R}_2-48\,\mathcal{R}_6)h\,\Box\Pi_{\mu\nu}\,\Pi^{\mu\rho}\Pi^\nu_{\;\;\rho}
\nonumber\\
&
-16\,\left(
\mathcal{R}_3
\,\Pi^{\mu\nu}\Pi-\mathcal{R}_2\,\Pi^{\mu}_{\;\;\rho}\Pi^{\rho\nu}
\right)
\left(h_{\mu\nu}\,\Box\Pi-2\,h_{\mu\sigma}\Box\Pi^\sigma_{\;\;\nu}\right)
-2\left(\mathcal{R}_1+16\,\mathcal{R}_5\right)\partial_\mu\partial_\nu\Pi_{\rho\sigma}\,\Pi^{\rho\sigma}\left(h\,\Pi^{\mu\nu}-2\,h^{\mu}_{\;\;\alpha}\Pi^{\alpha\nu}\right)
\nonumber\\
&
-16\,(\mathcal{R}_3+4\,\mathcal{R}_4-2\,\mathcal{R}_5)h\,\Box\Pi_{\mu\nu}\,\Pi^{\mu\nu}\Pi+h_{\mu\nu}\partial^\mu\partial^\nu\Pi^{\rho\sigma}\left[
32\,(2\,\mathcal{R}_4-\mathcal{R}_5)\Pi_{\rho\sigma}\Pi -2\,(\mathcal{R}_1-48\,\mathcal{R}_6)\Pi_{\rho\alpha}\Pi^\alpha_{\;\;\sigma}
\right]
\nonumber\\
&
-2\,h_{\mu\nu}\Box\Pi_{\rho\sigma}\left(
16 \,\mathcal{R}_5 \,\Pi^{\mu\nu}\Pi^{\rho\sigma}+\mathcal{R}_1 \,\Pi^{\mu\rho}\Pi^{\nu\sigma}
\right)
\nonumber\\
&
+h\,\partial_\mu\Pi\left[
16\,(\mathcal{R}_2+\mathcal{R}_3-2\,\mathcal{R}_7)
\partial^\mu\,\Pi\,\Pi-4\,(\mathcal{R}_1+8\,\mathcal{R}_2+32\,\mathcal{R}_4)\partial^\mu\Pi_{\rho\sigma}\Pi^{\rho\sigma}
-2\,(\mathcal{R}_1-16\mathcal{R}_2+8\,\mathcal{R}_3)\partial_\nu\Pi\,\Pi^{\mu\nu}
\right]
\nonumber\\
&
+h\,\partial_\mu\Pi_{\rho\sigma}\left[
-16\,(\mathcal{R}_2+4\mathcal{R}_4-2\,\mathcal{R}_5) \partial^\mu\Pi^{\rho\sigma}\Pi
+2(3\,\mathcal{R}_1-16\,\mathcal{R}_5-96\,\mathcal{R}_6)\partial_\nu\Pi^{\rho\sigma}\Pi^{\mu\nu}
\right]
\nonumber\\
&
+h_{\mu\nu}\partial^\rho\Pi^{\mu\nu}\left[
-32\,\mathcal{R}_2 \,\partial_\rho\Pi\,\Pi
+4\,\mathcal{R}_1 \,\partial_{\rho}\Pi_{\alpha\beta}\Pi^{\alpha\beta}
+4(\mathcal{R}_1-8\,\mathcal{R}_2)\partial^\sigma\Pi\,\Pi_{\rho\sigma}
\right]
\nonumber\\
&
+h_{\mu\nu}\partial^\mu\Pi\left[
-16\,(\mathcal{R}_3-2\,\mathcal{R}_7)\partial^\nu\Pi\,\Pi
+32\,(\mathcal{R}_2+4\,\mathcal{R}_4)\partial^\nu\Pi^{\rho\sigma}\,\Pi_{\rho\sigma}
-32\,(\mathcal{R}_2-\mathcal{R}_3)\partial_\rho\Pi\,\Pi^{\nu\rho}
\right]
\nonumber\\
&
+h_{\mu\nu}\partial^\mu\Pi^{\rho\sigma}\left[
32\,(\mathcal{R}_2+2\,\mathcal{R}_4-\mathcal{R}_5)\partial^\nu\Pi_{\rho\sigma}\,\Pi
-8(\mathcal{R}_1-24\,\mathcal{R}_6)\partial^\nu\Pi_{\rho\alpha}\,\Pi^{\alpha}_{\;\;\sigma}
\right.\nonumber\\
&\qquad\qquad\qquad\left.-4(\mathcal{R}_1-16\,\mathcal{R}_5)\partial_\alpha\Pi_{\rho\sigma}\,\Pi^{\nu\alpha}
+4(\mathcal{R}_1+8\,\mathcal{R}_2)\partial_\rho\Pi\,\Pi^\nu_{\;\;\sigma}
\right]
\nonumber\\
&
-16\,h_{\mu\nu}\Pi^{\mu\nu}\left(\mathcal{R}_3 \,\partial_\rho\Pi\,\partial^\rho\Pi +2\,\mathcal{R}_5\,\partial_\alpha\Pi_{\rho\sigma}\partial^\alpha\Pi^{\rho\sigma}\right)
\,,
\end{align}
where
\begin{equation}
\mathcal{R}_1 \equiv 
\frac{\mathcal{A}_1^2+\mathcal{A}_1-8\,\mathcal{A}_4}{\mathcal{A}_1^3}\,,
\quad
\mathcal{R}_2 \equiv 
\frac{\mathcal{B}_4}{\mathcal{A}_1^3}\,,
\quad
\mathcal{R}_3 \equiv 
\frac{\mathcal{C}_2}{\mathcal{A}_1^3}\,,
\quad
\mathcal{R}_4 \equiv 
\frac{\mathcal{C}_3}{\mathcal{A}_1^3}\,,
\quad
\mathcal{R}_5 \equiv 
\frac{\mathcal{B}_5}{\mathcal{A}_1^3}\,,
\quad
\mathcal{R}_6 \equiv 
\frac{\mathcal{B}_6}{\mathcal{A}_1^3}\,,
\quad
\mathcal{R}_7 \equiv 
\frac{\mathcal{D}_1}{\mathcal{A}_1^3}\,,
\end{equation}
Since we have already added boundary terms to isolate $h_{\mu\nu}$ without any derivatives, all of these terms contain high derivatives and cannot be further eliminated. To simultaneously remove all of these terms, we need $\mathcal{R}_1=\mathcal{R}_2=\mathcal{R}_3=\mathcal{R}_4=\mathcal{R}_5=\mathcal{R}_6=\mathcal{R}_7=0$, or,
\begin{equation}
\mathcal{A}_4 = \frac{\mathcal{A}_1(\mathcal{A}_1+1)}{8}\,,\quad
\mathcal{B}_4=\mathcal{B}_5=\mathcal{B}_6=\mathcal{C}_2=\mathcal{C}_3=\mathcal{D}_1=0\,.
\label{eq:noLambda4}
\end{equation}

To summarise, combining the conditions for consistent background \eqref{eq:background-consistency}, avoiding ghost modes in the free theory \eqref{eq:noquadraticghost}, removing the $\Lambda_5$ cubic interactions \eqref{eq:noLambda5} and $\Lambda_4$ quartic interactions \eqref{eq:noLambda4}, we obtain the following conditions:
\begin{gather}
\mathcal{A}_2 = 1\,,
\quad
\mathcal{A}_3=\frac{\mathcal{A}_1(\mathcal{A}_1-3)}{4}\,,
\quad
\mathcal{A}_4 = \frac{\mathcal{A}_1(\mathcal{A}_1+1)}{8}\,,
\label{coefA}\\
\mathcal{B}_1 = \mathcal{B}_2 = \mathcal{B}_3 = \mathcal{B}_4 = \mathcal{B}_5 = \mathcal{B}_6 = 0\,,\\
\mathcal{C}_1 = \mathcal{C}_2 = \mathcal{C}_3 = 0\,,\\
\mathcal{D}_1 =0\,,
\label{coefD}
\end{gather}
i.e. we obtain three conditions on the value of the coefficients on the background, six conditions on their first derivatives, three conditions on second derivatives and one condition on the third derivative. In principle, this procedure can be extended to higher order interactions. However, at quartic order, we observe that only the quantity $\mathcal{A}_1$ survives the stability conditions below $\Lambda_3$ at quartic order. Remarkably, the functional form of the coefficients turned out to be irrelevant: all four coefficients $C$, $D$, $E$ and $F$ are forced to be constant up to quartic order. We will consider the constant coefficients as a separate case in the following section.

\section{Disformal relation with constant coefficients}
\label{sec:composite}

The conditions for stability below $\Lambda_3$ obtained in Sec.\ref{sec:decouplinglimit} indicate that the coefficients $C$, $D$, $E$ and $F$ in \eqref{eq:mostgeneraltilde} should be constant, at least up to quartic order in perturbations around flat space-time. In this section, we investigate the special case of constant coefficients. 

For this case, the only relevant combinations are $\mathcal{A}_n$ which are constrained by conditions \eqref{coefA}. Using their definitions given in Appendix \ref{app:coefficients}, these conditions fix two of the coefficients
\begin{align}
D= \pm 2\,\sqrt{C}\,\sqrt{E}\,,\qquad F=0\,,
\end{align}
while the normalisation of the background metric in the Jordan frame can be determined by solving
\begin{equation}
\left(\sqrt{C}\sqrt{\Omega_0}\pm\sqrt{E}\right)^2=1\,.
\end{equation}
With these restrictions, the relation between the dRGT and Jordan frame metrics \eqref{eq:mostgeneraltilde} simply becomes
\begin{equation}
\tilde{g}_{\mu\nu} = C\,g_{\mu\nu} \pm 2\,\sqrt{C}\,\sqrt{E}g_{\mu\sigma}\,(\sqrt{\gamma})^{\sigma}_{\;\;\nu}+E\,f_{\mu\nu}\,,
\end{equation}
or in matrix form
\begin{equation}
\tilde{g} = C\, g \left(\mathbb{1}\pm\frac{\sqrt{E}}{\sqrt{C}}\sqrt{\gamma}\right)^2\,.
\label{eq:constantcoeffs}
\end{equation}
Left-multiplying both sides by $f^{-1}$, we get
\begin{equation}
\tilde{\gamma}^{-1} = C\,\left(\gamma^{-1/2}\pm \frac{\sqrt{E}}{\sqrt{C}}\,\mathbb{1}\right)^2\,.
\end{equation}
Taking the matrix square-root of both sides, we can invert this relation 
\begin{equation}
\gamma^{-1} = \frac{1}{C}\left(\tilde{\gamma}^{-1/2}+\frac{\sqrt{E}}{\sqrt{C}}\,\mathbb{1}\right)^2\,.
\end{equation}
Finally, by contracting with $f$ from the left, we can rewrite this relation in the component form as
\begin{equation}
g_{\mu\nu} = \alpha^2\,\tilde{g}_{\mu\nu} + 2\,\alpha\,\beta
\,\tilde{g}_{\mu\sigma}\sqrt{\tilde{\gamma}}^{\sigma}_{\;\;\nu} + \beta^2\,f_{\mu\nu}\,,
\label{eq:composite}
\end{equation}
where we defined
\begin{equation}
\alpha \equiv \frac{1}{\sqrt{C}}\,,\qquad
\beta \equiv \pm \frac{\sqrt{E}}{\sqrt{C}}\,.
\end{equation}
Thus the matter couples minimally to metric $g_{\mu\nu}$, which is related to the dRGT frame metric $\tilde{g}_{\mu\nu}$ disformally with constant coefficients. This is nothing but the composite matter coupling introduced in \cite{deRham:2014naa}. The stability conditions \eqref{coefA} imply that this coupling is the unique disformal coupling with constant coefficients that does not generate the Boulware-Deser mode below $\Lambda_3$. Our perturbative study in the previous section reveals that any functional dependence in the coefficients are forbidden at least up to quartic order.

There is a simpler way to see why this example actually works. Let us consider only the longitudinal scalar field interactions around flat space-time, yet in a non-perturbative manner. In this case, we have 
\begin{equation}
g=\eta\,,\qquad \gamma= (1+\partial\partial\pi)^2\,,
\end{equation}
where $\partial\partial\pi$ denotes the Hessian matrix for $\pi$. We can then rearrange \eqref{eq:constantcoeffs} to get
\begin{equation}
\tilde{g} = \left[
(\sqrt{C}\pm\sqrt{E})\mathbb{1}\pm \sqrt{E}\,\partial\partial\pi\right]\eta
\left[
(\sqrt{C}\pm\sqrt{E})\mathbb{1}\pm \sqrt{E}\,\partial\partial\pi\right]\,,
\end{equation}
or, in component form we can rewrite it as
\begin{equation}
\tilde{g}_{\mu\nu} = \eta_{\rho\sigma}\,\frac{\partial y^{\rho}}{\partial x^\mu}\,\frac{\partial y^{\sigma}}{\partial x^\nu}\,,
\end{equation}
where
\begin{equation}
y^{\mu} \equiv (\sqrt{C}\pm\sqrt{E})x^\mu \pm \sqrt{E}\,\partial^\mu\pi\,.
\end{equation}
This form of the transformation reveals why the composite coupling is special. The  transformation is simply a non-linear coordinate transformation for the flat metric, thus keeps the Einstein-Hilbert action invariant. Although we did not allow the space-time perturbations, we introduced non-linear St\"uckelberg perturbations. The composite coupling, thanks to its full-squared form, prevents the generation of the most dangerous derivative interactions. 

\section{Discussion}
\label{sec:discussion}
In this paper, we explored the possibility of new interactions by keeping the matter coupling generic. We introduced a new geometry relevant for matter dynamics, which is related to the original geometry via a four-field generalisation of the disformal relation. In dRGT massive gravity preserving the number of degrees of freedom is not possible for general matter couplings. Instead, we extracted information on what our four free functions should be, by requiring the new degree of freedom does not appear at least within the strong coupling scale $\Lambda_3$. Perturbatively, we calculated dangerous interaction terms and obtained stability conditions, which revealed that all four functions need to be constant up to quartic order interactions.

The constant coefficient case provides a unique relation between the dRGT and Jordan frames, which coincides with the composite metric coupling scenario, proposed in Ref.~\cite{deRham:2014naa} to control quantum corrections from spoiling the dRGT tuning. Our approach gives a new perspective for this coupling. The cosmology of massive gravity with composite coupling is known to evade the cosmological no-go result \cite{deRham:2014naa, Gumrukcuoglu:2014xba,Solomon:2014iwa} and has a stable de Sitter attractor \cite{Heisenberg:2016spl}, which is in agreement with background observations \cite{Heisenberg:2016dkj}. In a scenario where the standard model follows the geodesics of the composite metric, it is expected that the propagation speed of gravitational waves will generically be different that the speed of light and will be constrained by the observation of gravitational and electromagnetic wave  from neutron start merger GW170817 \cite{Monitor:2017mdv}. Such a constraint is available for the bimetric version of the composite coupling \cite{Akrami:2018yjz} although similar constraints may affect the massive gravity case at hand. On the other hand, a stable cosmology only requires a single sector to couple compositely. Therefore, a scenario which breaks the weak equivalence principle, where the Standard model couples minimally to one metric, while a hidden sector that couples compositely would be unconstrained by the gravitational wave propagation bounds.

Due to the perturbative nature of our study, we were not able to provide a proof at non-linear level for the uniqueness of the composite metric coupling.  Therefore we cannot conclude whether viable examples other than the composite coupling exist. The total-squared form of the composite case allows it to evade dangerous high derivative interactions that stem from the Einstein-Hilbert term which would be very difficult to emulate when the four functions are not constant.  However, the relation between different geometries that we used is built out of the tensor $\sqrt{\partial_\mu\phi^a \partial_\nu\phi^b}$, which naturally reproduces the composite coupling. Conversely, a naive generalisation of disformal relations to four functions built out of $\partial_\mu\phi^a \partial_\nu\phi^b$ would yield non-trivial values for the derivatives of the coefficients, and to recover the composite case we would have to remove all low energy interactions at all orders in perturbation theory. We therefore cannot exclude the possibility that another convenient parameterisation to yield another example that cannot be uncovered in the current approach.

Our formulation can easily be extended to the case where the translation symmetry of the scalar fields are broken. This would allow us to introduce field dependencies in the four functions. Although this could complicate the perturbative study, removing the field derivative dependence can potentially reveal new interactions in the so-called ``generalised massive gravity'' proposed by Ref.\cite{deRham:2014gla}.

Finally, if one uses an external field to define a new geometry for matter (as opposed to the four St\"uckelberg fields) the survival of the dRGT constraint is less of an issue, although this increases the number of degrees of freedom  with respect to the original dRGT massive gravity \cite{Wittner:2018jrq,Golovnev:2018icm}.

\acknowledgments
We thank Clare Burrage, Matteo Fasiello, Rampei Kimura, Ryo Namba, Atsushi Naruko, Tony Padilla and Gianmassimo Tasinato for illuminating discussions. The research leading to these results has received funding from the European Research Council
(ERC) under the European Union’s Horizon 2020 research and innovation programme (grant agreement 646702 ”CosTesGrav”). KK is supported by the UK STFC grant ST/N000668/1.

\appendix 
\section{Form of the generalised disformal transformation}
\label{app:proofofdisformal}
Here we provide a derivation of the disformal transformation \eqref{eq:mostgeneraltilde} in the fashion of Bekenstein \cite{Bekenstein:1992pj}. We write the Finsler line element as
\begin{equation}
ds^2 = g_{\alpha\beta}dx^\alpha dx^\beta F(\{I_A\},\{H_A\})\,,
\end{equation}
where with the Poincar\'e symmetry of the field space in mind, we did not consider direct dependence on the fields $\phi^a$. For $ds^2$ to be a homogeneous function of second order in $dx^\alpha$, we choose $I_A$ and $H_A$ as
\begin{equation}
I_A \equiv [X^A]\,,\qquad
H_A \equiv \frac{g_{\mu\rho}(X^A)^\rho_{\;\;\nu}dx^\mu dx^\nu}{-g_{\alpha\beta}dx^\alpha dx^\beta}\,,
\end{equation}
where $X^\mu_{\;\;\alpha} X^\alpha_{\;\;\nu}= g^{\mu\alpha}f_{\alpha\nu}$ and $A$ denotes the order of matrix power. \footnote{One can also define the functions $H_A$ and $I_A$ without the square root, by replacing $X$ with $g^{-1}f$, resulting in transformations \eqref{eq:nosquareroot-disformal}. Although the formulation as equivalent (see Appendix \ref{app:squarerootformalism}) we adopt the square-root formulation in the main text as it is better suited to the form of the dRGT action.}
Since $X^4$ can be written in terms of the lower powers of $X$ and characteristic polynomials $e_n(X)$, $H_A$ runs through $A=1,2,3$ while $I_A$ runs through $A=1,2,3,4$. The combinations $I_A$ and $H_A$ are thus the only quantities of this form that are degree-zero homogeneous functions of $dx^\alpha$ that depend on the first derivatives of the fields $\phi^a$. 

Given these definitions, we can extract the quasimetric $\tilde{g}_{\mu\nu}$ via \cite{Bekenstein:1992pj}
\begin{equation}
\tilde{g}_{\mu\nu} = \frac{1}{2} \frac{\partial^2 (ds^2)}{\partial dx^\mu\,\partial dx^\nu}\,,
\end{equation}
which gives
\begin{align}
 \tilde{g}_{\mu\nu} =& \left(F-\frac{\partial F}{\partial H_A}\,H_A\right)g_{\mu\nu} - \frac{\partial F}{\partial H_A}\,g_{\mu\alpha}(X^A)^\alpha_{\;\;\nu}
 \nonumber\\
 &-\frac{2\,g_{\alpha\tau}g_{\beta\eta}dx^\alpha dx^\beta}{-g_{\rho\sigma}dx^\rho dx^\sigma}\frac{\partial^2 F}{\partial H_A\partial H_B}\,\left[
 (X^A)^\tau_{\;\;\mu}(X^B)^\eta_{\;\;\nu}+(X^A)^\tau_{\;\;\mu} H_B\delta^\eta_\nu+(X^B)^\tau_{\;\;\nu} H_A\delta^\eta_\mu+H_AH_B\delta^\tau_\mu\delta^\eta_\nu\right]\,,
\end{align}
where summation convention also applies to upper case latin indices. In order to obtain a Riemannian geometry, we impose that $\tilde{g}_{\mu\nu}$ is independent of $dx^\alpha$. This is achieved  if  the second derivative of $F$ vanishes.\footnote{In the single scalar field case of Ref.\cite{Bekenstein:1992pj}, the first term in the square brackets is also independent of $dx^\alpha$, although in this general case, this is not true for all values of $A$ and $B$.} Thus the Finsler factor reduces to the form
\begin{equation}
F = C(\{I_B\}) + D_A(\{I_B\}) H_A\,.
\end{equation}
Using this expression, the quasimetric then becomes:
\begin{equation}
\tilde{g}_{\mu\nu} = C(\{[X^B]\}) g_{\mu\nu} - D_A(\{[X^B]\}) g_{\mu\alpha}(X^A)^\alpha_{\;\;\nu}\,,
\end{equation}
which we can rewrite as
\begin{equation}
\tilde{g}_{\mu\nu} = C(\{[X^A]\}) g_{\mu\nu} + D(\{[X^A]\}) g_{\mu\alpha}X^\alpha_{\;\;\nu} + E(\{[X^A]\}) f_{\mu\nu} + F (\{[X^A]\}) f_{\mu\alpha}X^\alpha_{\;\;\nu}\,,
\end{equation}
which is precisely the form of Eq.\eqref{eq:mostgeneraltilde}.

\section{Equivalence of the formulations}
\label{app:squarerootformalism}
In this Appendix, we show that the formulation with square-roots \eqref{eq:mostgeneraltilde} is equivalent to the disformal transformations of four scalar fields \eqref{eq:nosquareroot-disformal}.

We start with the Cayley-Hamilton theorem for $4\times4$ for square matrices:
\begin{equation}
\gamma^2 = e_1(\sqrt{\gamma})\gamma^{3/2}-e_2(\sqrt{\gamma})\gamma+e_3(\sqrt{\gamma})\sqrt{\gamma}-e_4(\sqrt{\gamma})\mathbb{1}\,.
\label{eq:sqrtidentity}
\end{equation}
By multiplying this relation twice with $\sqrt{\gamma}$ and using the intermediate relation, we obtain the following:
\begin{equation}
\gamma^3= (e_3-2\,e_1e_2+e_1^3)\gamma^{3/2}-(e_4-e_1e_3-e_2^2+e_1^2e_2)\gamma+(e_1^2e_3-e_2 e_3-e_1e_4)\gamma^{1/2}-e_4(e_1^2-e_2)\mathbb{1}\,,
\label{eq:3halves}
\end{equation}
where $e_n=e_n(\sqrt{\gamma})$. We can now use the above two equations to solve for $\sqrt{\gamma}$ and $\gamma^{3/2}$ and write them in terms of $\gamma$, $\gamma^2$ and $\gamma^3$. This is the first step in building a relation between the two formulations. The missing piece is to relate the characteristic polynomials of the two formulations $e_n(\sqrt{\gamma})$ and $E_n(\gamma)$. By multiplying \eqref{eq:3halves} twice with $\sqrt{\gamma}$ and replacing all the half integer powers of $\gamma$, we obtain
\begin{equation}
\gamma^4= (e_1^2-2\,e_2)\gamma^3-(2\,e_4-2\,e_1e_3+e_2^2)\gamma^2+(e_3^2-2\,e_2e_4)\gamma-e_4^2\mathbb{1}\,.
\end{equation}
On the other hand, we also have the analogue of Eq.\eqref{eq:sqrtidentity} for $\gamma$ which is
\begin{equation}
\gamma^4 = E_1(\gamma)\gamma^{3}-E_2(\gamma)\gamma^2+E_3(\gamma)\gamma-E_4(\gamma)\mathbb{1}\,.
\end{equation}
The above two relations allow us to relate the characteristic polynomials for $\gamma$ and $\sqrt{\gamma}$. It is convenient to rewrite the relation in terms of traces:
\begin{align}
\left(3 \left([\gamma]^2-2 [\gamma^2]\right)-6 [\gamma] [\sqrt{\gamma}]^2+[\sqrt{\gamma}]^4+8 [\sqrt{\gamma}] [\gamma^{3/2}]\right)^2=24 \left([\gamma]^4-6 [\gamma]^2 [\gamma^2]+8 [\gamma] [\gamma^3]+3 [\gamma^2]^2-6
   [\gamma^4]\right)\,,\nonumber\\
3 [\gamma]^3+9 [\sqrt{\gamma}]^2 \left([\gamma]^2-2 [\gamma^2]\right)-18 [\gamma] [\gamma^2]-9 [\gamma] [\sqrt{\gamma}]^4+24 [\gamma^3]+[\sqrt{\gamma}]^6+16 [\sqrt{\gamma}]^3 [\gamma^{3/2}]-8 [\gamma^{3/2}]^2=0\,.
\end{align}
These two equations give $[\sqrt{\gamma}]$ and $[\gamma^{3/2}]$ in terms of $[\gamma]$, $[\gamma^2]$, $[\gamma^3]$ and $[\gamma^4]$. This shows that formulation with $\sqrt{\gamma}$ is equivalent to the one with $\gamma$. 

On the other hand, the system above has $16$ solutions \footnote{The first equation has two solution for $[\gamma^{3/2}]$ which is a polynomial in $[\sqrt\gamma]$. Using each of these solutions in the second equation yields a polynomial equation for $[\sqrt{\gamma}]$ of eighth order, hence there are 16 solutions.}, so solving it is equivalent to taking the square-root of a $4\times4$ matrix. In other words, given a theory in $\gamma$ formulation, there are $16$ distinct equivalent $\sqrt{\gamma}$ formulations. However, when considering perturbations around flat space, $\gamma$ is proportional to identity. Thus, taking the square-root amounts to choosing the sign of each diagonal entry. Once the background is fixed to, say, $\sqrt{\gamma}=\mathbb{1}$, then the complicated non-linear system of algebraic equations above become a very simple linear system for the perturbations.

To summarise, the two formulations are equivalent, but starting from $\gamma$ formalism, there is no single way to generically go to $\sqrt{\gamma}$ formulation. However, around a simple background with $\gamma=\mathbb{1}$, fixing $\sqrt{\gamma}=\mathbb{1}$ picks a single solution and we argue that the decoupling limit analysis to be equivalent in both languages.

\section{Writing the exact theory in the Jordan frame}
\label{app:jordan}
In this section, we consider the most general disformal transformation as defined in Eq.\eqref{eq:mostgeneraltilde}, using the $\sqrt{\gamma}$ tensor as the building block. This relation allows us to express the dRGT frame metric $\tilde{g}$ in terms of the Jordan frame metric $g$. We first start by relating the determinants of the two metrics as
\begin{align}
\frac{\det\tilde{g}}{\det{g}}=&\,C^4+C^3\,\left[D\,e_1+\left(e_1^2-2\,e_2\right)\,E+\left(e_1^3-3\,e_2\,e_1+3\,e_3\right)\,F\right]
\nonumber\\
&
+C^2\,\left[D^2\,e_2+D\,\left(e_1\,e_2-3\,e_3\right)\,E+D\,\left(e_2\,e_1^2-e_3\,e_1-2\,e_2^2+4\,e_4\right)
\,F+\left(e_2^2-2\,e_1\,e_3+2\,e_4\right)\,E^2
\right.\nonumber\\
&
\left.\qquad\qquad+\left(-2\,e_3\,e_1^2+e_2^2\,e_1+5\,e_4\,e_1-e_2\,e_3\right)\,E\,F+\left(e_2^3-3\,e_1\,e_3\,e_2-3\,e_4\,e_2+3\,e_3^2+3\,e_1^2\,e_4\right)\,F^2\right]
\nonumber\\
&+C\,\left[D^3\,e_3+D^2
\,\left(e_1\,e_3-4\,e_4\right)\,E+D^2\,\left(e_3\,e_1^2-e_4\,e_1-2\,e_2\,e_3\right)\,F+D\,\left(e_2\,e_3-3\,e_1\,e_4\right)\,E^2
\right.\nonumber\\
&\qquad\qquad
+D\,\left(-3\,e_4\,e_1^2+e_2\,e_3\,e_1-3\,e_3^2+4\,e_2\,e_4\right)\,E\,F+D\,\left(e_3
\,e_2^2-e_1\,e_4\,e_2-2\,e_1\,e_3^2+5\,e_3\,e_4\right)\,F^2
\nonumber\\
&\qquad\qquad+\left(e_3^2-2\,e_2\,e_4\right)\,E^3+\left(e_1\,e_3^2-e_4\,e_3-2\,e_1\,e_2\,e_4\right)\,E^2\,F+\left(-2\,e_4\,e_2^2+e_3^2\,e_2+4\,e_4^2-e_1\,e_3\,e_4\right)
\,E\,F^2
\nonumber\\
&\left.\qquad\qquad+\left(e_3^3-3\,e_2\,e_4\,e_3+3\,e_1\,e_4^2\right)\,F^3\right]
\nonumber\\
&+D^4\,e_4+D^3\,e_1\,e_4\,E+D^3\,\left(e_1^2-2\,e_2\right)\,e_4\,F+D^2\,e_2\,e_4\,E^2
+D^2\,\left(e_1\,e_2-3\,e_3\right)\,e_4\,E\,F
\nonumber\\
&
+D^2
\,e_4\,\left(e_2^2-2\,e_1\,e_3+2\,e_4\right)\,F^2+D\,e_3\,e_4\,E^3+D\,\left(e_1\,e_3-4\,e_4\right)\,e_4\,E^2\,F
+D\,e_4\,\left(e_2\,e_3-3\,e_1\,e_4\right)\,E\,F^2
\nonumber\\
&+D\,e_4\,\left(e_3^2-2\,e_2\,e_4\right)\,F^3+e_4^2
\,E^4+e_1\,e_4^2\,E^3\,F+e_2\,e_4^2\,E^2\,F^2
+e_3\,e_4^2\,E\,F^3+e_4^3\,F^4\,,
\label{eq:dettilde}
\end{align}
where the elementary polynomials $e_n=e_n(\sqrt{\gamma})$ are given in \eqref{eq:elementarypoly}. With this, we can express the inverse metric in the following form
\begin{equation}
\tilde{g}^{\mu\nu} = \frac{\det{g}}{\det{\tilde{g}}}\left[\tilde{C}([\sqrt{\gamma}^n])\,g^{\mu\nu} + \tilde{D}([\sqrt{\gamma}^n])\, (\sqrt{\gamma}\,g^{-1})^{\mu\nu} + \tilde{E}([\sqrt{\gamma}^n])\, (\gamma \,g^{-1})^{\mu\nu} + \tilde{F}([\sqrt{\gamma}^n]) (\gamma^{3/2}\,g^{-1})^{\mu\nu}\right]\,,
\end{equation}
where the coefficients are given by
\begin{align}
\tilde{C}([\sqrt{\gamma}^n])
&\equiv
\left[C^3+D\,e_1
\,C^2+E\,\left(e_1^2-2\,e_2\right)\,C^2+F\,\left(e_1^3-3\,e_2\,e_1+3\,e_3\right)\,C^2+D^2\,e_2\,C+D\,E\,\left(e_1\,e_2-3\,e_3\right)\,C
\right.\nonumber\\
&\qquad
+E^2\,\left(e_2^2-2\,e_1\,e_3+e_4\right)\,C+D\,F\,\left(e_2\,e_1^2-e_3\,e_1-2
\,e_2^2+2\,e_4\right)\,C
\nonumber\\
&\qquad
+E\,F\,\left(-2\,e_3\,e_1^2+e_2^2\,e_1+3\,e_4\,e_1-e_2\,e_3\right)\,C+F^2\,\left(e_2^3-3\,e_1\,e_3\,e_2-2\,e_4\,e_2+3\,e_3^2+2\,e_1^2\,e_4\right)\,C+D^3\,e_3
\nonumber\\
&\qquad
+D^2\,F\,\left(e_1^2-2\,e_2\right)\,e_3+D^2
\,E\,\left(e_1\,e_3-e_4\right)+D\,F^2\,e_3\,\left(e_2^2-2\,e_1\,e_3+2\,e_4\right)+D\,E^2\,\left(e_2\,e_3-e_1\,e_4\right)
\nonumber\\
&\qquad
+E^3\,\left(e_3^2-e_2\,e_4\right)+D\,E\,F\,\left(-e_4\,e_1^2+e_2\,e_3\,e_1-3
\,e_3^2+2\,e_2\,e_4\right)+F^3\,\left(e_3^3-2\,e_2\,e_4\,e_3+e_1\,e_4^2\right)
\nonumber\\
&\left.\qquad
+E\,F^2\,\left(-e_4\,e_2^2+e_3^2\,e_2+e_4\,\left(e_4-e_1\,e_3\right)\right)+E^2\,F\,\left(e_1\,\left(e_3^2-e_2\,e_4\right)-e_3
\,e_4\right)\right]\,,\nonumber\\
\tilde{D}([\sqrt{\gamma}^n])
&\equiv-e_2\,D^3-C\,e_1\,D^2-E\,e_1\,e_2\,D^2+F
\,\left(-e_2\,e_1^2+2\,e_2^2-e_4\right)\,D^2-C^2\,D+C\,E\,\left(2\,e_2-e_1^2\right)\,D
\nonumber\\
&\qquad
+C\,F\,\left(-e_1^3+3\,e_2\,e_1-e_3\right)\,D+E^2\,\left(e_4-e_2^2\right)\,D+F^2\,\left(-e_2^3+2\,e_1\,e_3\,e_2-e_1^2\,e_4\right)
\,D
\nonumber\\
&\qquad
+E\,F\,\left(3\,e_2\,e_3-e_1\,\left(e_2^2+e_4\right)\right)\,D+C\,E^2\,e_3+C\,E\,F\,\left(2\,e_1\,e_3-2\,e_4\right)+C\,F^2\,\left(e_3\,e_1^2-e_4\,e_1-e_2\,e_3\right)
\nonumber\\
&\qquad
+E^3\,\left(e_1\,e_4-e_2
\,e_3\right)+E\,F^2\,e_2\,\left(2\,e_1\,e_4-e_2\,e_3\right)+E^2\,F\,\left(e_4\,e_1^2-e_2\,e_3\,e_1+e_2\,e_4\right)
\nonumber\\
&\qquad
+F^3\,\left(e_4\,e_2^2-e_3^2\,e_2-e_4^2+e_1\,e_3\,e_4\right)\,,\nonumber\\
\tilde{E}([\sqrt{\gamma}^n])
&\equiv e_1\,D^3+E\,e_1^2\,D^2+C\,D^2+F\,\left(e_1^3-2\,e_1\,e_2\right)\,D^2-C\,E\,e_1\,D-2\,C\,F\,e_2\,D+E^2\,e_1\,e_2\,D
\nonumber\\
&\qquad
+E\,F\,\left(e_2\,e_1^2-3\,e_3\,e_1+2\,e_4\right)\,D+F^2\,e_1\,\left(e_2^2-2\,e_1\,e_3+2
\,e_4\right)\,D-C^2\,E+C\,E^2\,\left(e_2-e_1^2\right)
\nonumber\\
&\qquad+C\,E\,F\,\left(-e_1^3+e_2\,e_1-e_3\right)+E^2\,F\,e_1\,\left(e_1\,e_3-2\,e_4\right)+E^3\,\left(e_1\,e_3-e_4\right)+C\,F^2\,\left(-e_2
\,e_1^2+e_3\,e_1+e_2^2-e_4\right)
\nonumber\\
&\qquad+E\,F^2\,\left(-e_4\,e_1^2+e_2\,e_3\,e_1-e_2\,e_4\right)+F^3\,\left(e_1\,e_3^2-e_4\,e_3-e_1\,e_2\,e_4\right)\,,\nonumber\\
\tilde{F}([\sqrt{\gamma}^n])
&\equiv-D^3-E\,e_1\,D^2+F\,\left(2\,e_2-e_1^2\right)\,D^2+2\,C\,E\,D+C\,F\,e_1\,D-E^2\,e_2\,D+E\,F\,\left(3\,e_3-e_1\,e_2\right)\,D
\nonumber\\
&\qquad
+F^2\,\left(-e_2^2+2\,e_1\,e_3-e_4\right)\,D+C\,E\,F\,e_1^2-C^2\,F+C
\,E^2\,e_1+C\,F^2\,\left(e_1\,e_2-2\,e_3\right)-E^3\,e_3
\nonumber\\
&\qquad
+E^2\,F\,\left(e_4-e_1\,e_3\right)+E\,F^2\,\left(e_1\,e_4-e_2\,e_3\right)+F^3\,\left(e_2\,e_4-e_3^2\right)\,.
\end{align}
Using these, the Christoffel symbols for the $\tilde{g}$ metric can be obtained as
\begin{equation}
\tilde{\Gamma}^{\alpha}_{\beta\gamma} = \Gamma^{\alpha}_{\beta\gamma} + \Delta\Gamma^{\alpha}_{\beta\gamma}\,,
\end{equation}
where 
\begin{equation}
\Delta\Gamma^{\alpha}_{\beta\gamma} \equiv\frac{\tilde{g}^{\alpha\delta}}{2}\left(\nabla_\beta \tilde{g}_{\delta\gamma}+\nabla_\gamma \tilde{g}_{\beta\delta}-\nabla_\delta \tilde{g}_{\beta\gamma}\right)\,,
\end{equation}
and $\nabla$ is the covariant derivative operator compatible with $g$. Finally, the Ricci scalar for the tilde metric is computed via
\begin{equation}
\tilde{R} = R_{\mu\nu}\tilde{g}^{\mu\nu} + \tilde{g}^{\mu\nu}\tilde{g}_{\beta\rho}\tilde{g}^{\alpha\gamma}\left(\Delta\Gamma^\rho_{\nu\gamma}\Delta\Gamma^\beta_{\alpha\mu}-
\Delta\Gamma^\rho_{\alpha\gamma}\Delta\Gamma^\beta_{\mu\nu}\right) +\tilde{g}^{\alpha\rho}\tilde{g}^{\mu\nu}\left(\nabla_\alpha\nabla_\nu\tilde{g}_{\mu\rho}-\nabla_\mu\nabla_\nu\tilde{g}_{\alpha\rho}\right)\,.
\end{equation}
This expression, along with the determinant \eqref{eq:dettilde} allows us to write the kinetic part of the action \eqref{eq:action} in the Jordan frame. Generically, this introduces a derivative coupling between the $g$ and $f$ metrics through $\nabla\tilde{g}$ type terms. 

For the mass terms in Eq.\eqref{eq:action}, our options are limited. These depend on the tensor $\sqrt{\tilde{\gamma}}=\sqrt{\tilde{g}^{-1}f}$, where technically the square-root operation cannot be performed in an exact manner. For these terms, we will rely on the existence of a vacuum solution where one can unambiguously evaluate the square-roots, then discuss the mass terms perturbatively.

\section{Form of the coefficients in the perturbative expansion}
\label{app:coefficients}
In this section, we define some combinations of the derivatives of coefficients $C$--$F$ in \eqref{eq:mostgeneraltilde} evaluated on the Minkowski background. 

We first define
\begin{align}
C_0 & \equiv C\vert_{g/\Omega_0=f=\eta}\,,\nonumber\\
C_i &\equiv \frac{\partial C}{\partial [\sqrt{\gamma}^i]}\Big\vert_{g/\Omega_0=f=\eta}\,,\nonumber\\
C_{ij} &\equiv \frac{\partial^2 C}{\partial [\sqrt{\gamma}^i]\partial [\sqrt{\gamma}^j]}\Big\vert_{g/\Omega_0=f=\eta}\,,\nonumber\\
C_{ijk}&\equiv \frac{\partial^3 C}{\partial [\sqrt{\gamma}^i]\partial [\sqrt{\gamma}^j]\partial [\sqrt{\gamma}^k]}\Big\vert_{g/\Omega_0=f=\eta}\,,
\end{align}
 and similar relations hold for the functions $D$, $E$ and $F$. Using these we define the quantities $\mathcal{A}$ which contain only the background values of the coefficients, while $\mathcal{B}$, $\mathcal{C}$ and $\mathcal{D}$ depend only on the first, second and third derivatives, respectively. The combinations below exhaust all coefficients that appear in the expansion of the action up to quartic order in perturbations:
\begin{align}
\mathcal{A}_1\equiv&-\frac{\sqrt{\Omega_0} D_0}{2}-C_0 \Omega_0+\frac{F_0}{2
   \sqrt{\Omega_0}}\,,\nonumber\\
\mathcal{A}_2\equiv&\sqrt{\Omega_0} D_0+E_0+C_0
   \Omega_0+\frac{F_0}{\sqrt{\Omega_0}}\,,\nonumber\\
\mathcal{A}_3\equiv&\frac{3 \sqrt{\Omega_0} D_0}{8}+C_0
   \Omega_0-\frac{F_0}{8 \sqrt{\Omega_0}}\,,\nonumber\\
\mathcal{A}_4\equiv&\frac{5 F_0}{16 \sqrt{\Omega_0}}-\frac{D_0
   \sqrt{\Omega_0}}{16}\,,\nonumber\\
\mathcal{B}_1\equiv&C_2+\frac{3 C_3}{\sqrt{\Omega_0}}+\frac{6
   C_4}{\Omega_0}+\frac{D_2}{\sqrt{\Omega_0}}+\frac{3 D_3}{\Omega_0}+\frac{6
   D_4}{\Omega_0^{3/2}}+\frac{E_2}{\Omega_0}+\frac{3 E_3}{\Omega_0^{3/2}}+\frac{6
   E_4}{\Omega_0^2}+\frac{F_2}{\Omega_0^{3/2}}+\frac{3 F_3}{\Omega_0^2}+\frac{6
   F_4}{\Omega_0^{5/2}}\,,\nonumber\\
\mathcal{B}_2\equiv&\frac{3 \sqrt{\Omega_0} C_1}{8}+C_2+\frac{15 C_3}{8
   \sqrt{\Omega_0}}+\frac{3 C_4}{\Omega_0}+\frac{3 D_1}{8}+\frac{D_2}{\sqrt{\Omega_0}}+\frac{15
   D_3}{8 \Omega_0}+\frac{3 D_4}{\Omega_0^{3/2}}+\frac{3 E_1}{8
   \sqrt{\Omega_0}}+\frac{E_2}{\Omega_0}+\frac{15 E_3}{8 \Omega_0^{3/2}}+\frac{3
   E_4}{\Omega_0^2}
   \nonumber\\
   &+\frac{3 F_1}{8 \Omega_0}+\frac{F_2}{\Omega_0^{3/2}}+\frac{15 F_3}{8
   \Omega_0^2}+\frac{3 F_4}{\Omega_0^{5/2}}\,,\nonumber\\
\mathcal{B}_3\equiv&\frac{3 \sqrt{\Omega_0} C_1}{2}+3 C_2+\frac{9
   C_3}{2 \sqrt{\Omega_0}}+\frac{6 C_4}{\Omega_0}+\frac{5 D_1}{4}+\frac{5 D_2}{2
   \sqrt{\Omega_0}}+\frac{15 D_3}{4 \Omega_0}+\frac{5
   D_4}{\Omega_0^{3/2}}+\frac{E_1}{\sqrt{\Omega_0}}+\frac{2 E_2}{\Omega_0}+\frac{3
   E_3}{\Omega_0^{3/2}}+\frac{4 E_4}{\Omega_0^2}
   \nonumber\\
   &+\frac{3 F_1}{4 \Omega_0}+\frac{3 F_2}{2
   \Omega_0^{3/2}}+\frac{9 F_3}{4 \Omega_0^2}+\frac{3
   F_4}{\Omega_0^{5/2}}\,,\nonumber\\
\mathcal{B}_4\equiv&-\frac{D_1}{16}-\frac{D_2}{8 \sqrt{\Omega_0}}-\frac{3 D_3}{16
   \Omega_0}-\frac{D_4}{4 \Omega_0^{3/2}}+\frac{3 F_1}{16 \Omega_0}+\frac{3 F_2}{8
   \Omega_0^{3/2}}+\frac{9 F_3}{16 \Omega_0^2}+\frac{3 F_4}{4 \Omega_0^{5/2}}\,,\nonumber\\
\mathcal{B}_5\equiv&\frac{3
   \sqrt{\Omega_0} C_1}{8}+C_2+\frac{15 C_3}{8 \sqrt{\Omega_0}}+\frac{3 C_4}{\Omega_0}+\frac{3
   D_1}{16}+\frac{D_2}{2 \sqrt{\Omega_0}}+\frac{15 D_3}{16 \Omega_0}+\frac{3 D_4}{2
   \Omega_0^{3/2}}-\frac{3 F_1}{16 \Omega_0}-\frac{F_2}{2 \Omega_0^{3/2}}-\frac{15 F_3}{16
   \Omega_0^2}-\frac{3 F_4}{2 \Omega_0^{5/2}}\,,\nonumber\\
\mathcal{B}_6\equiv&\frac{5 \sqrt{\Omega_0} C_1}{16}+C_2+\frac{35
   C_3}{16 \sqrt{\Omega_0}}+\frac{4 C_4}{\Omega_0}+\frac{5
   D_1}{16}+\frac{D_2}{\sqrt{\Omega_0}}+\frac{35 D_3}{16 \Omega_0}+\frac{4
   D_4}{\Omega_0^{3/2}}+\frac{5 E_1}{16 \sqrt{\Omega_0}}+\frac{E_2}{\Omega_0}+\frac{35 E_3}{16
   \Omega_0^{3/2}}+\frac{4 E_4}{\Omega_0^2}
   \nonumber\\
   &+\frac{5 F_1}{16
   \Omega_0}+\frac{F_2}{\Omega_0^{3/2}}+\frac{35 F_3}{16 \Omega_0^2}+\frac{4
   F_4}{\Omega_0^{5/2}}\,,\nonumber\\
\mathcal{C}_1\equiv&\frac{C_{11}}{4}+\frac{C_{12}}{\sqrt{\Omega_0}}+\frac{3
   C_{13}}{2 \Omega_0}+\frac{2 C_{14}}{\Omega_0^{3/2}}+\frac{C_{22}}{\Omega_0}+\frac{3
   C_{23}}{\Omega_0^{3/2}}+\frac{4 C_{24}}{\Omega_0^2}+\frac{9 C_{33}}{4 \Omega_0^2}+\frac{6
   C_{34}}{\Omega_0^{5/2}}+\frac{4 C_{44}}{\Omega_0^3}+\frac{D_{11}}{4
   \sqrt{\Omega_0}}+\frac{D_{12}}{\Omega_0}+\frac{3 D_{13}}{2 \Omega_0^{3/2}}
   \nonumber\\
   &+\frac{2
   D_{14}}{\Omega_0^2}+\frac{D_{22}}{\Omega_0^{3/2}}+\frac{3 D_{23}}{\Omega_0^2}+\frac{4
   D_{24}}{\Omega_0^{5/2}}+\frac{9 D_{33}}{4 \Omega_0^{5/2}}+\frac{6 D_{34}}{\Omega_0^3}+\frac{4
   D_{44}}{\Omega_0^{7/2}}+\frac{E_{11}}{4 \Omega_0}+\frac{E_{12}}{\Omega_0^{3/2}}+\frac{3
   E_{13}}{2 \Omega_0^2}+\frac{2 E_{14}}{\Omega_0^{5/2}}+\frac{E_{22}}{\Omega_0^2}
   \nonumber\\
   &+\frac{3
   E_{23}}{\Omega_0^{5/2}}+\frac{4 E_{24}}{\Omega_0^3}+\frac{9 E_{33}}{4 \Omega_0^3}+\frac{6
   E_{34}}{\Omega_0^{7/2}}+\frac{4 E_{44}}{\Omega_0^4}+\frac{F_{11}}{4
   \Omega_0^{3/2}}+\frac{F_{12}}{\Omega_0^2}+\frac{3 F_{13}}{2 \Omega_0^{5/2}}+\frac{2
   F_{14}}{\Omega_0^3}+\frac{F_{22}}{\Omega_0^{5/2}}+\frac{3 F_{23}}{\Omega_0^3}+\frac{4
   F_{24}}{\Omega_0^{7/2}}
   \nonumber\\
   &+\frac{9 F_{33}}{4 \Omega_0^{7/2}}+\frac{6 F_{34}}{\Omega_0^4}+\frac{4
   F_{44}}{\Omega_0^{9/2}}\,,\nonumber\\
\mathcal{C}_2\equiv&\frac{C_{11}}{4}+\frac{C_{12}}{\sqrt{\Omega_0}}+\frac{3
   C_{13}}{2 \Omega_0}+\frac{2 C_{14}}{\Omega_0^{3/2}}+\frac{C_{22}}{\Omega_0}+\frac{3
   C_{23}}{\Omega_0^{3/2}}+\frac{4 C_{24}}{\Omega_0^2}+\frac{9 C_{33}}{4 \Omega_0^2}+\frac{6
   C_{34}}{\Omega_0^{5/2}}+\frac{4 C_{44}}{\Omega_0^3}+\frac{D_{11}}{8
   \sqrt{\Omega_0}}+\frac{D_{12}}{2 \Omega_0}+\frac{3 D_{13}}{4
   \Omega_0^{3/2}}
   \nonumber\\
   &+\frac{D_{14}}{\Omega_0^2}+\frac{D_{22}}{2 \Omega_0^{3/2}}+\frac{3 D_{23}}{2
   \Omega_0^2}+\frac{2 D_{24}}{\Omega_0^{5/2}}+\frac{9 D_{33}}{8 \Omega_0^{5/2}}+\frac{3
   D_{34}}{\Omega_0^3}+\frac{2 D_{44}}{\Omega_0^{7/2}}-\frac{F_{11}}{8
   \Omega_0^{3/2}}-\frac{F_{12}}{2 \Omega_0^2}-\frac{3 F_{13}}{4
   \Omega_0^{5/2}}-\frac{F_{14}}{\Omega_0^3}-\frac{F_{22}}{2 \Omega_0^{5/2}}
   \nonumber\\
   &-\frac{3 F_{23}}{2
   \Omega_0^3}-\frac{2 F_{24}}{\Omega_0^{7/2}}-\frac{9 F_{33}}{8 \Omega_0^{7/2}}-\frac{3
   F_{34}}{\Omega_0^4}-\frac{2 F_{44}}{\Omega_0^{9/2}}\,,\nonumber\\
\mathcal{C}_3\equiv&\frac{3 C_{11}}{16}+\frac{7
   C_{12}}{8 \sqrt{\Omega_0}}+\frac{3 C_{13}}{2 \Omega_0}+\frac{9 C_{14}}{4
   \Omega_0^{3/2}}+\frac{C_{22}}{\Omega_0}+\frac{27 C_{23}}{8 \Omega_0^{3/2}}+\frac{5
   C_{24}}{\Omega_0^2}+\frac{45 C_{33}}{16 \Omega_0^2}+\frac{33 C_{34}}{4 \Omega_0^{5/2}}+\frac{6
   C_{44}}{\Omega_0^3}+\frac{3 D_{11}}{16 \sqrt{\Omega_0}}+\frac{7 D_{12}}{8 \Omega_0}
   \nonumber\\
   &+\frac{3
   D_{13}}{2 \Omega_0^{3/2}}+\frac{9 D_{14}}{4 \Omega_0^2}+\frac{D_{22}}{\Omega_0^{3/2}}+\frac{27
   D_{23}}{8 \Omega_0^2}+\frac{5 D_{24}}{\Omega_0^{5/2}}+\frac{45 D_{33}}{16 \Omega_0^{5/2}}+\frac{33
   D_{34}}{4 \Omega_0^3}+\frac{6 D_{44}}{\Omega_0^{7/2}}+\frac{3 E_{11}}{16 \Omega_0}+\frac{7
   E_{12}}{8 \Omega_0^{3/2}}+\frac{3 E_{13}}{2 \Omega_0^2}
   \nonumber\\
   &+\frac{9 E_{14}}{4
   \Omega_0^{5/2}}+\frac{E_{22}}{\Omega_0^2}+\frac{27 E_{23}}{8 \Omega_0^{5/2}}+\frac{5
   E_{24}}{\Omega_0^3}+\frac{45 E_{33}}{16 \Omega_0^3}+\frac{33 E_{34}}{4 \Omega_0^{7/2}}+\frac{6
   E_{44}}{\Omega_0^4}+\frac{3 F_{11}}{16 \Omega_0^{3/2}}+\frac{7 F_{12}}{8 \Omega_0^2}+\frac{3
   F_{13}}{2 \Omega_0^{5/2}}+\frac{9 F_{14}}{4 \Omega_0^3}
   \nonumber\\
   &+\frac{F_{22}}{\Omega_0^{5/2}}+\frac{27
   F_{23}}{8 \Omega_0^3}+\frac{5 F_{24}}{\Omega_0^{7/2}}+\frac{45 F_{33}}{16 \Omega_0^{7/2}}+\frac{33
   F_{34}}{4 \Omega_0^4}+\frac{6 F_{44}}{\Omega_0^{9/2}}\,,\nonumber\\
\mathcal{D}_1\equiv&\frac{C_{111}}{8
   \sqrt{\Omega_0}}+\frac{3 C_{112}}{4 \Omega_0}+\frac{9 C_{113}}{8 \Omega_0^{3/2}}+\frac{3 C_{114}}{2
   \Omega_0^2}+\frac{3 C_{122}}{2 \Omega_0^{3/2}}+\frac{9 C_{123}}{2 \Omega_0^2}+\frac{6
   C_{124}}{\Omega_0^{5/2}}+\frac{27 C_{133}}{8 \Omega_0^{5/2}}+\frac{9 C_{134}}{\Omega_0^3}+\frac{6
   C_{144}}{\Omega_0^{7/2}}+\frac{C_{222}}{\Omega_0^2}
   \nonumber\\
   &+\frac{9 C_{223}}{2 \Omega_0^{5/2}}+\frac{6
   C_{224}}{\Omega_0^3}+\frac{27 C_{233}}{4 \Omega_0^3}+\frac{18 C_{234}}{\Omega_0^{7/2}}+\frac{12
   C_{244}}{\Omega_0^4}+\frac{27 C_{333}}{8 \Omega_0^{7/2}}+\frac{27 C_{334}}{2 \Omega_0^4}+\frac{18
   C_{344}}{\Omega_0^{9/2}}+\frac{8 C_{444}}{\Omega_0^5}+\frac{D_{111}}{8 \Omega_0}
   \nonumber\\
   &+\frac{3
   D_{112}}{4 \Omega_0^{3/2}}+\frac{9 D_{113}}{8 \Omega_0^2}+\frac{3 D_{114}}{2 \Omega_0^{5/2}}+\frac{3
   D_{122}}{2 \Omega_0^2}+\frac{9 D_{123}}{2 \Omega_0^{5/2}}+\frac{6 D_{124}}{\Omega_0^3}+\frac{27
   D_{133}}{8 \Omega_0^3}+\frac{9 D_{134}}{\Omega_0^{7/2}}+\frac{6
   D_{144}}{\Omega_0^4}+\frac{D_{222}}{\Omega_0^{5/2}}
   \nonumber\\
   &+\frac{9 D_{223}}{2 \Omega_0^3}+\frac{6
   D_{224}}{\Omega_0^{7/2}}+\frac{27 D_{233}}{4 \Omega_0^{7/2}}+\frac{18 D_{234}}{\Omega_0^4}+\frac{12
   D_{244}}{\Omega_0^{9/2}}+\frac{27 D_{333}}{8 \Omega_0^4}+\frac{27 D_{334}}{2
   \Omega_0^{9/2}}+\frac{18 D_{344}}{\Omega_0^5}+\frac{8 D_{444}}{\Omega_0^{11/2}}+\frac{E_{111}}{8
   \Omega_0^{3/2}}
   \nonumber\\
   &+\frac{3 E_{112}}{4 \Omega_0^2}+\frac{9 E_{113}}{8 \Omega_0^{5/2}}+\frac{3 E_{114}}{2
   \Omega_0^3}+\frac{3 E_{122}}{2 \Omega_0^{5/2}}+\frac{9 E_{123}}{2 \Omega_0^3}+\frac{6
   E_{124}}{\Omega_0^{7/2}}+\frac{27 E_{133}}{8 \Omega_0^{7/2}}+\frac{9 E_{134}}{\Omega_0^4}+\frac{6
   E_{144}}{\Omega_0^{9/2}}+\frac{E_{222}}{\Omega_0^3}+\frac{9 E_{223}}{2 \Omega_0^{7/2}}
   \nonumber\\
   &+\frac{6
   E_{224}}{\Omega_0^4}+\frac{27 E_{233}}{4 \Omega_0^4}+\frac{18 E_{234}}{\Omega_0^{9/2}}+\frac{12
   E_{244}}{\Omega_0^5}+\frac{27 E_{333}}{8 \Omega_0^{9/2}}+\frac{27 E_{334}}{2 \Omega_0^5}+\frac{18
   E_{344}}{\Omega_0^{11/2}}+\frac{8 E_{444}}{\Omega_0^6}+\frac{F_{111}}{8 \Omega_0^2}+\frac{3
   F_{112}}{4 \Omega_0^{5/2}}+\frac{9 F_{113}}{8 \Omega_0^3}
   \nonumber\\
   &+\frac{3 F_{114}}{2 \Omega_0^{7/2}}+\frac{3
   F_{122}}{2 \Omega_0^3}+\frac{9 F_{123}}{2 \Omega_0^{7/2}}+\frac{6 F_{124}}{\Omega_0^4}+\frac{27
   F_{133}}{8 \Omega_0^4}+\frac{9 F_{134}}{\Omega_0^{9/2}}+\frac{6
   F_{144}}{\Omega_0^5}+\frac{F_{222}}{\Omega_0^{7/2}}+\frac{9 F_{223}}{2 \Omega_0^4}+\frac{6
   F_{224}}{\Omega_0^{9/2}}+\frac{27 F_{233}}{4 \Omega_0^{9/2}}
   \nonumber\\
   &+\frac{18 F_{234}}{\Omega_0^5}+\frac{12
   F_{244}}{\Omega_0^{11/2}}+\frac{27 F_{333}}{8 \Omega_0^5}+\frac{27 F_{334}}{2
   \Omega_0^{11/2}}+\frac{18 F_{344}}{\Omega_0^6}+\frac{8 F_{444}}{\Omega_0^{13/2}}\,.
\end{align}

\end{document}